\documentclass[a4paper,11pt]{article}
\pdfoutput=1 

\usepackage{jcappub}
\usepackage[utf8]{inputenc}
\usepackage[T1]{fontenc}

\usepackage{multicol}
\usepackage{placeins}

\usepackage{subcaption}
\usepackage{siunitx}

\usepackage{soul}

\newcommand{\HI}{H\,{\sc i}}
\newcommand{\Lya}{Lyman-$\alpha$}
\newcommand{\dv}{$\Delta\nu$}

\newcommand{\comment}[1]{}

\usepackage{bm}

\title{Lyman-$\alpha$ polarization from cosmological ionization fronts: I. Radiative transfer simulations}

\author[a,b,c]{Yuanyuan Yang,}
\author[a,d]{Emily Koivu,}
\author[a,d]{Chenxiao Zeng,}
\author[a,d]{Heyang Long,}
\author[a,b,d]{and Christopher M. Hirata}
\affiliation[a]{Center for Cosmology and AstroParticle Physics, The Ohio State University, 191 West Woodruff Avenue, Columbus, OH 43210, USA}
\affiliation[b]{Department of Astronomy, The Ohio State University, 140 West 18th Avenue, Columbus, OH 43210, USA}
\affiliation[c]{Khoury College of Computer Science, Northeastern University, 440 Huntington Ave, Boston, MA 02115, USA}
\affiliation[d]{Department of Physics, The Ohio State University, 191 West Woodruff Avenue, Columbus, OH 43210, USA}
\emailAdd{yang.4904@buckeyemail.osu.edu}

\abstract{In this paper, we present the formalism of simulating \Lya\ emission and polarization around reionization ($z$ = 8) from a plane-parallel ionization front. We accomplish this by using a Monte Carlo method to simulate the production of a \Lya\ photon, its propagation through an ionization front, and the eventual escape of this photon. This paper focuses on the relation of the input parameters of ionization front speed $U$, blackbody temperature $T_{\rm bb}$, and neutral hydrogen density $n_{\rm HI}$, on intensity $I$ and polarized intensity $P$ as seen by a distant observer. The resulting values of intensity range from $3.18\times 10^{-14}$ erg/cm$^{2}$/s/sr to $1.96 \times 10^{-9}$ erg/cm$^{2}$/s/sr , and the polarized intensity ranges from $5.73\times 10^{-17}$ erg/cm$^{2}$/s/sr to $5.31 \times 10^{-12}$ erg/cm$^{2}$/s/sr. We found that higher $T_{\rm bb}$, higher $U$, and higher $n_{\rm HI}$ contribute to higher intensity, as well as polarized intensity, though the strongest dependence was on the hydrogen density. The dependence of viewing angle of the front is also explored. We present tests to support the validity model, which makes the model suitable for further use in a following paper where we will calculate the intensity and polarized intensity power spectrum on a full reionization simulation.}


\date{\today}

\begin{document}
\maketitle
\flushbottom

\section{Introduction}

Following the Big Bang, the universe expanded and cooled. At the time of recombination, the temperature dropped low enough for the hydrogen gas to transition from ionized to neutral. Later, when the first stars and galaxies formed, the ultraviolet radiation they produced led the hydrogen in the intergalactic medium to become reionized. The epoch of reionization is of broad interest for both cosmologists and astrophysicists. From the cosmologist's perspective, the ionization and thermal history of the Universe is key to studies of novel sources of energy injection \cite{2016MNRAS.460.1885U} and dark matter physics \cite{2017PhRvD..96b3522I}; it underlies studies of the power spectrum with the \Lya\ forest \cite{1998MNRAS.296...44G, 2019MNRAS.487.1047M}; and through its effect on the normalization of cosmic microwave background (CMB) anisotropies it affects studies of dark energy and modified gravity \cite{2016PhRvD..93d3013L}. From the astrophysicist's perspective, reionization is interesting as a way of learning about the early sources of ionizing radiation \cite{2015ApJ...802L..19R, 2015ApJ...810...71F, 2019ApJ...879...36F}, in its own right as a major event in the history of intergalactic matter, and through its feedback effect on the formation of small galaxies \cite{2000ApJ...539..517B}.

We have several ways of probing the history and structure of reionization, and intensity mapping (IM) is one of the novel methods. Different from a traditional galaxy survey, intensity mapping collects and statistically analyses the emission line from galaxies and intergalactic medium, which can give information at larger scales and higher redshifts \cite{2017arXiv170909066K}. Intensity mapping complements the other major probes of the reionization epoch. The large-scale polarization of the CMB due to Thomson scattering after reionization gives a global constraint and suggests that the midpoint of reionization occurred at redshift $z\approx 7.7\pm0.7$ \cite{2020A&A...641A...6P}. \Lya\ absorption gives constraints on individual lines of sight, and is sensitive to neutral gas; even a small amount of \HI\ can result in a deep absorption trough in the spectrum of a quasar or other source \cite{1965ApJ...142.1633G}. Such troughs are observed at $z\gtrsim 6$ \cite{2001AJ....122.2833F, 2001AJ....122.2850B}, although the interpretation is complicated by the modest number of sightlines and the fact that \Lya\ absorption saturates at even a small ($\sim 10^{-4}$) neutral fraction \cite{2006ARA&A..44..415F}. One can also probe reionization using \Lya\ emitters (LAEs) since neutral gas in the IGM can scatter \Lya\ photons out of the line of sight; thus LAEs can provide insights into the history and structure of reionization \cite{2004ApJ...617L...5M, 2004MNRAS.349.1137S, 2017ApJ...842L..22Z, 2017ApJ...844...85O, 2018PASJ...70S..16K}. In the near future, it may be possible to map neutral hydrogen during the reionization epoch using \HI\ 21 cm emission \cite{1997ApJ...475..429M, 1999A&A...345..380S}. Current and future experiments for the 21-cm mapping include the Canadian Hydrogen Intensity Mapping Experiment
(CHIME)\footnote{http://chime.phas.ubc.ca/}, the Five hundred meter Aperture Spherical Telescope (FAST)\footnote{http://fast.bao.ac.cn/en/}, and the
Square Kilometer Array (SKA)\footnote{https://www.skatelescope.org/}. This is a faint line, and the foreground challenges in this part of the radio spectrum are significant, but the 21 cm line is optically thin at IGM densities and it does not require a background source.

One of the ways we hope to learn about reionization is through the \Lya\ intensity mapping of the epoch of reionization \cite{2013ApJ...763..132S, 2014ApJ...786..111P}. \Lya\ is the spectral line of neutral hydrogen corresponding to the 1s--2p transition, with a wavelength of 1216 $\si{\angstrom}$. The \Lya\ line also has immense diagnostic power because when a \Lya\ photon encounters a hydrogen atom, it scatters instead of being destroyed.
Because \Lya\ photons scatter off of even small column densities of \HI, we might expect it to be linearly polarized, and for this polarization to encode important geometrical information since the most likely direction of polarization is perpendicular to the plane of the last scattering. \Lya\ polarization has been studied as a probe of galaxies and their environments \cite{2018ApJ...856..156E}. Prior to reionization, we expect a source to be surrounded by a ``halo'' of scattered, polarized \Lya\ light \cite{1999ApJ...524..527L, 2022arXiv221209630C}; this may be modified depending on the local velocity structure of the \HI\ around the source \cite{2008MNRAS.386..492D}. There is one detailed study so far of the polarization power spectrum of \Lya\ in the reionization epoch \cite{2020PhRvD.101h3032M}, which focused on scattered radiation from galaxies. On larger scales, however, we might expect the ionization fronts themselves to contribute significantly to the polarized intensity mapping signal. Cosmological ionization fronts are warm and partially ionized, so they should cool by \Lya\ emission from collisionally excited \HI\ \cite{1994MNRAS.266..343M, 2008ApJ...672...48C, 2016MNRAS.457.3006D}, and due to multiple scattering, this radiation should be polarized when it emerges from the front. The direction of polarization should be related to the orientation of the ionization front. Moreover, ionization bubbles are coherent over large scales (likely tens of cMpc), and thus are a good candidate for contributing to the power spectrum at large scales. Detection of this signal with future instruments could in principle be an interesting diagnostic of the geometrical structure of reionization.

In this paper, we will create a model for simulating \Lya\ emissions from reionization in a single plane-parallel ionization front. This includes generating, propagating, and scattering \Lya\ photon processes, and will predict the final status, especially the intensity and polarization of \Lya\ radiation. In the companion paper (``Paper II''), we will to extend this to a cosmological simulated ionization front.

\section{Ionization front models}
\label{sec:formalism}

In this work, we make use of the gird model in \cite{2021ApJ...906..124Z} to parameterize the characteristics of ionization fronts (I-front). In the fiducial model of \cite{2021ApJ...906..124Z}, the I-front has velocity $U = 5\times 10^8$ cm $\rm s^{-1}$ and incident radiation with blackbody temperature $T_{\rm bb} = 5\times 10^4$ K. The 1-dimensional I-front is split into $N_{\rm grid} = 2000$ cells,  each cell with physical width $N_{\rm H}/n_{\rm H}$ (unit: cm), where $N_{\rm H}$ (unit: $\rm cm^{-2}$) is the column density of hydrogen (\HI\ + H\,{\sc ii}) and $n_{\rm H}$ (unit: $\rm cm^{-3}$) is the total number density of hydrogen.

In this work, we extend the parameter space of I-fronts for the sake of general investigation of various I-fronts related physics scenarios. We simulate I-fronts with speed $U\in\left[5\times 10^7 ,\,5 \times 10^9 \right]\, \rm cm\, s^{-1}$, incident radiation blackbody temperature $ T_{\rm bb}\in \left[5\times 10^4,\, 10^5\right]\,\rm K$, and \HI\ density $n_{\rm HI}\in \left[1.37\times 10^{-5},\,1.37\times 10^{-3} \right]\,\rm cm^{-3}$ (i.e. 0.1 to 10 times \HI\ mean density at $z = 8$). By covering this parameter space, we anticipate that we will encompass the astrophysical sources of reionization; the typical blackbody temperatures for reionization would fall into this temperature range, and 5$\times$10$^{8}$ cm/s is a typical cosmological front speed. We also center our density range around the cosmic mean, and allow it to vary from 10 times less dense to 10 times denser than the mean. 

We find that in our simulations, it is possible for a photon in the damping wings of the Lyman-$\alpha$ line to travel far into the neutral side of the front, and then re-scatter. To achieve convergence of this effect, we set $N_{\rm grid}=2\times 10^5$ (with slabs of column density $\Delta N_{\rm H}=2.5\times 10^{16}\,{\rm cm}^{-2}$).

We modify the initial temperature while using the simulation suite in \cite{2021ApJ...906..124Z} out of numerical complication. For thermal evolution of the ionization front, it is sufficient to start from idealized ``cold'' initial conditions with temperature near zero. However, for radiative transfer calculations, the lower the temperature, the narrower the Doppler width, which leads to quite short length scales for a photon to redshift through a line,
which is challenging for our numerical method. For this paper, we used the temperature $ T_{\rm i} = 1.75$ K  based on standard cosmological recombination and Compton de-coupling \cite{2011PhRvD..83d3513A}, with no additional heating sources such as X-ray heating.

We use the cosmological model from {\slshape Planck} 2018 results \cite{2020A&A...641A...6P}: $H_{\rm 0} = 67.4 \,\rm km\, s^{-1} Mpc^{-1}$, $\Omega_{\rm m} = 0.315$ and $\Omega_{\rm b}\rm h^2= 0.0224$. 

\section{Method: Monte Carlo}
\label{sec:MC}

With the I-front model set, we could simulate the production, propagation, and escape of \Lya\ photons through it using the method of Monte Carlo. We track several properties of the photon:
\begin{itemize}
    \item The position $Z$, which ranges from 0 (ionized side) to $N_{\rm grid}-1$ (neutral side). 
    \item The frequency offset $\Delta\nu = \nu-\nu_{{\rm Ly}\alpha}$ (measured so that $\Delta\nu>0$ on the blue side of the line and $\Delta\nu<0$ on the red side of the line). 
    \item The direction of propagation $\mu=\cos\theta$ (measured so that $\mu=+1$ points to the neutral side and $\mu=-1$ points to the ionized side).
    \item The linear polarization $p=Q/I$ (measured so that $p=+1$ is polarized in the North-South plane containing the $z$-direction, and $p=-1$ is polarized in the East-West plane perpendicular to the $z$-direction).
\end{itemize}
Since the ionization front model is plane-parallel, we do not need to track the $x$ or $y$ coordinate of the photon or the longitude angle $\phi$ of the propagation direction. Also, the diagonal ($U/I$) and circular ($V/I$) polarizations of the photon are zero due to symmetry.

\begin{figure}
\centering
\includegraphics[scale=0.65]{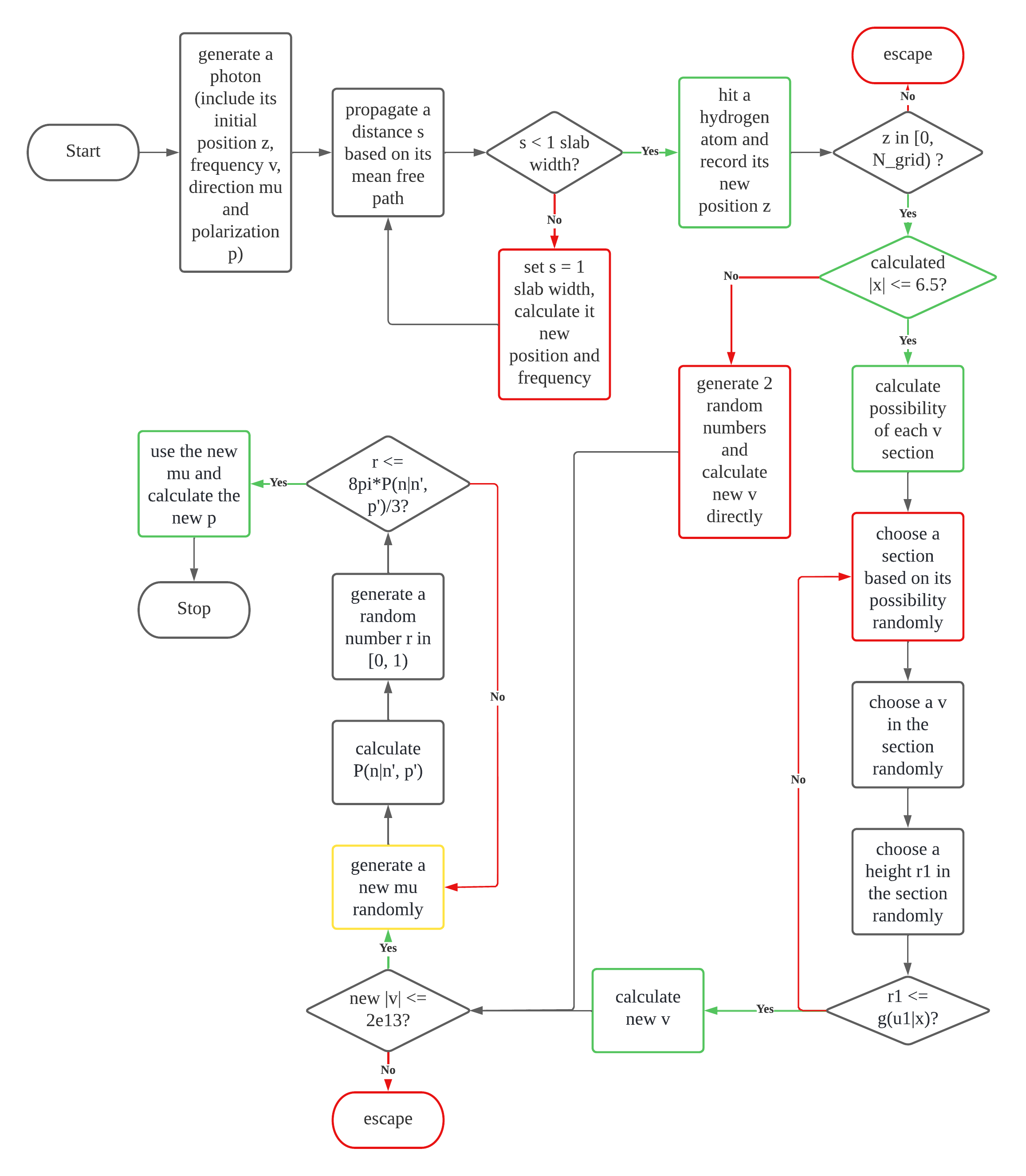}
\caption{\large Flowchart of the Monte Carlo method \label{fig:FC}}
\centering
\end{figure}

To begin the simulation of each photon, we place it at a position drawn from the emissivity distribution, give it an isotropic initial direction and polarization, and set its initial frequency based on the Voigt profile. Then we propagate the photon based on its mean free path until it hits a hydrogen atom and scatters, or escapes from the simulation grid. We describe the details of generating a photon in \S\ref{sec:GP}, propagation in \S\ref{sec:PP}, and scattering in \S\ref{sec:SP}. Figure $\ref{fig:FC}$ is the flowchart of the whole Monte Carlo process.

\subsection{Generate a Photon}
\label{sec:GP}

Our simulation pipeline of \Lya\ photon propagation through the I-fronts starts with generating a \Lya\ photon with its initial position, frequency, direction, and polarization.

The probability distribution of the position $Z$ we place the photon is determined by the distribution of \Lya\ emissivity. The \Lya\ emissivity, which is the number of \Lya\ photons emitted per unit volume per unit time (unit: $\rm cm^{-3}$ $\rm s^{-1}$), is given by 
\begin{equation}
    \frac{dN}{dV\, dt} = n_{\rm e} n_{\rm HI} q_{{\rm Ly}\alpha}(T_{\rm e}),
    \label{eq:dN_dVdt}
\end{equation}
where $n_{\rm e}$ (unit: $\rm cm^{-3}$) is the electron density, $q_{{\rm Ly}\alpha} = q_{2p}+q_{3s}+q_{3d}$ (unit: $\rm cm^3$ $\rm s^{-1}$) is the total rate coefficient for hydrogen excitation that will result in Lyman-$\alpha$ emission 
\begin{equation}
{\rm H(1s)} + e^- ~\rightarrow~ {\rm H}(\rm nl) + e^-,
\end{equation}
and $T_{\rm e}$ is the electron temperature (unit: K). We use the rates tabulated in \cite{1983MNRAS.202P..15A}.

In our calculation, we consider the excitations of hydrogen from ground state to 2p, 3s, and 3d states; each of these could decay back to 1s with the emission of a Lyman-$\alpha$ photon. Other excitation channels either do not decay by emitting \Lya\ photons (e.g., 1s $\rightarrow$ 2s decays by emitting 2 continuum photons, 1s $\rightarrow 3p$ decays by 1  H$\alpha$ photon + 2 continuum photons) or the probabilities are small compared to the $n=2$ and 3 cases (i.e. $n\ge 4$ cases \cite{2006MNRAS.367..259H, 2006MNRAS.367.1057P}).

With the emission rate calculated in each slab, we could write cumulative probability for a photon to be generated in slab number $\le i$ as
\begin{equation}
P_{i}= \frac{\sum_{i'=0}^{i} (dN/dV\,dt)_{i'}}{\sum_{i'=0}^{N_{\rm grid}-1} (dN/dV\,dt)_{i'}}.
\end{equation}
Note that the total \Lya\ photons emissivity is converged within the I-front because: on the ionized side, the \HI\ density declines exponentially (atoms exposed to the incident UV flux are ionized); on the neutral side, the excitation rate $q_{{\rm Ly}\alpha}$ decays exponentially with the declination of temperature.

To determine the position of the photon, we draw a random number $\xi$ from uniform [0, 1] distribution and place the photon into slab $i$ if $P_{i-1}\le\xi<P_i$. Within each slab, we also randomly assign the exact position of the photon based on a uniform distribution at $i-1<Z<i$.

\begin{figure}
\centering
\includegraphics[width=75mm]{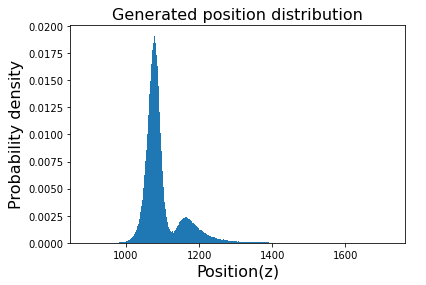}
\caption{Distribution of generated position}
\label{fig:posi}
\end{figure}

Figure \ref{fig:posi} shows the distribution of the generated position of photons. The initial position test runs $10^6$ photons under the setting of blackbody incident temperature $T_{\rm bb} = 5 \times 10^4$K, ionization front velocity $u = 5 \times 10^8 \rm cm$ $\rm s^{-1}$, and front hydrogen density $n_{\rm HI} = 1.37 \times 10^{-4}\, {\rm cm}^{-3}$. As the ionized side has no neutral hydrogen atom to excite and the neutral side has almost no electrons and is very cold, the \Lya\ emissivity should be zero on either side, as seen. In this figure, most of the \Lya\ photons emitted at slice number $Z\approx 1100$, where the ionization fraction of hydrogen first reaches 0.5, which matches to what we expected. The second peak is a result of $T_{\rm e}$ being larger than on the ionized side \cite{2021ApJ...906..124Z}; this is expected because of the exponential dependence of the excitation rate coefficients.

For the initial direction and polarization of photons, as the \Lya\ photons are emitted from hydrogens excited by isotropically distributed local thermalized electrons, we also implement an isotropic distribution of initial direction with $\mu=\cos\theta$ uniformly distributed between $-1$ and $+1$ and the initial polarization $p=0$. 

The last step is to assign frequency to the photons.  The frequency of \Lya\ emission obeys the Voigt distribution \cite{1938ApJ....88..508H}, which is
\begin{equation}
    \Phi(\Delta\nu) = \int_{-\infty}^\infty d\Delta\nu^\prime \Phi_{\rm G}(\Delta\nu^\prime) \Phi_{\rm L}(\Delta\nu - \Delta\nu^\prime)
\end{equation}
where $\Delta\nu = \nu-\nu_{{\rm Ly}\alpha}$ (unit: Hz) is the frequency offset from $\nu_{\rm Ly\alpha}$, and $\nu_{\rm Ly\alpha}$ (unit: Hz) is the center frequency of \Lya\ emission line. The Gaussian component due to thermal motion of the atoms is
\begin{equation}
\Phi_{\rm G}(\Delta\nu) = \frac1{\sqrt{2\pi}\,\sigma} e^{-\Delta\nu^2/(2\sigma^2)},
~~~
\sigma = \frac{\nu_{{\rm Ly}\alpha}\sqrt{kT_{\rm HI}/ m_{\rm HI} } }c,
\label{eq.G}
\end{equation}
and the Lorentzian component due to natural broadening of the 2p level is
\begin{equation}
\Phi_{\rm L}(\Delta\nu) = \frac{\gamma}{\pi(\gamma^2+\Delta\nu^2)},
~~~
\gamma = \frac1{4\pi T_{\rm 2p}}.
\end{equation}
where $k$ (unit: g $\rm cm^2$ $\rm s^{-2}$ $\rm K^{-1}$) is Boltzmann's constant, $T_{\rm HI}$ (unit: K) is the temperature of the neutral hydrogen atoms, $m_{\rm HI}$ (unit: g) is the mass of hydrogen atom, $C$ (unit: cm $\rm s^{-1}$) is the speed of light, and $T_{\rm 2p}$ (unit: s) is the lifetime of H(2p).

The total line profile $\Phi$ is a convolution of the two terms; the total frequency offset $\Delta\nu$ could be represented by the sum of frequency offset from the Gaussian kernel and the Lorenzian kernel, i.e. $\Delta\nu = \Delta\nu_{\rm G} +\Delta\nu_{\rm L}$.  We generate the Gaussian random variable $\Delta\nu_{\rm G}$ by the polar transformation method \cite{box58}, and the Lorentzian $\Delta\nu_{\rm L}$ by a univariate transformation method (since the inverse of the cumulative distribution function can be solved analytically).

\subsection{Propagate a Photon}
\label{sec:PP}

After assigning initial conditions to the \Lya\ photon, we could propagate it before it's absorbed by a hydrogen atom. We calculate the mean free path ($\ell_{\rm mfp}$, unit: cm) of the photon to quantify the propagation distance:
\begin{equation}
    \ell_{\rm mfp} = \frac{1}{n_{\rm HI} \sigma_{\rm tot}}
\end{equation}
where $\rm n_{\rm HI}$ is the number density of neutral hydrogen, which is calculated by initial hydrogen density times the neutral fraction of hydrogen in the slab. The total cross section $\sigma_{\rm tot}$ (unit: $\rm cm^{-2}$) generally counts all interactions, but in the frequency range  where only the Lyman-$\alpha$ transition (1s$\rightarrow$2p) is important, the cross section could be written as 
\begin{equation}
    \sigma_{\rm tot}(\Delta\nu) = \frac{ 3A\lambda_{{\rm Ly}\alpha}^2}{8\pi}\phi(\Delta\nu)
\end{equation}
where $A$ (unit: $\rm s^{-1}$) is the spontaneous decay rate for the transition from 2p to 1s, $\lambda_{{\rm Ly}\alpha}$ (unit: cm) is the wavelength of the \Lya\ photon, which will be eventually transferred into frequency, and $\phi(\Delta\nu)$ is the Voigt probability distribution.

The calculation of Voigt distribution is somewhat complicated, our calculation uses the series expansion in the Voigt parameter \cite{1948ApJ...108..112H} to second order ($H_2$ term). For the Dawson integral $F(x)$ that appears in the first order ($H_1$) term, we split it into two cases: the small $|x|$ case and the large $|x|$ case, where $x=\Delta\nu/\sqrt2\sigma$ is the frequency offset parameter. For $|x|<8$, we use the Taylor expansion, and for $|x|>8$ we use the asymptotic expansion as Taylor expansion overflows when $|x|>8$. The Taylor expansion of the Dawson integral is given by 
\begin{equation}
    F(x) = e^{-x^2}\sum_{n=0}^{x^2+10|x|}\frac{x^{2n+1}}{n!(2n+1)}
\end{equation}
and the asymptotic solution is given by 
\begin{equation}
    F(x) = \sum_{n=0}^{x^2 + 10|x|} 2^{-n-1} x^{2n-1} (2n-1)^{n} .
\end{equation}
The summation indicates the number of terms we have taken; the numerical calculation of $F(x)$ is at least accurate to 14 digit places at the matching point $|x| = 8$.

We generate the propagation distance $s$ of the photon from an exponential distribution of mean $\ell_{\rm mfp}$.
The relative distance of the photon to the moving ionization front is $\Delta s = s \mu - U s / c$. Finally, we divided the $\Delta \rm s$ by the width of the slab to convert it to our slab-based  system. The photon redshifting due to expansion of the Universe is expressed by
\begin{equation}
\Delta\nu_{\rm new} = \Delta\nu_{\rm old} 
- H(z) s \frac{\Delta \nu + \nu_{\rm Ly\alpha}} c,
\label{eq:dnn}
\end{equation}
where $H(z) = H_0 \sqrt{\Omega_m (1+z)^3 + \Omega_{\Lambda}}$ is the Hubble parameter (unit: $\rm s^{-1}$) at redshift $z$.

One complication is that in an inhomogeneous Universe, the mean free path $\ell_{\rm mfp}$ is not spatially uniform and the propagation distance is not exponentially distributed. We solve this by clipping the exponential distribution: if the photon enters the next slice, its position is reset to the boundary between slices, we compute the distance $s$ traveled before the photon reaches the boundary, and update the frequency according to Eq.~(\ref{eq:dnn}). We then re-propagate the photon in the same direction. The same clipping procedure is also applied if the photon propagates a distance $s$ larger than or equal to 1 slab width; this is especially important if the propagating direction is near parallel to the front, since then it may changes its frequency significantly before it hit a slab boundary.

\subsection{Scattering a Photon}
\label{sec:SP}

After propagating a certain distance, the \Lya\ photon could be absorbed and re-emitted by a hydrogen atom. In this section, we will elaborate the method of simulating this scattering process in order to get the post-scattering photon properties (i.e. frequency, direction and polarization).

After the \Lya\ photon moved a certain distance and was absorbed by a hydrogen atom, it will scatter rather than be destroyed and will have a new frequency, direction, and polarization. In this section, we will describe how to get those new properties step by step. This first involves solving for the velocity of the atom that causes the scattering, and then it involves drawing the outgoing direction (and hence frequency) and polarization in the I-front frame.

We use the rejection method based on \cite{1982ApJ...255..303L} to calculate the frequency and angular redistribution of the \Lya\ photon when both the Lorentzian broadening and Maxwellian velocity dispersion of the scatterers are important. Before getting started, we want to introduce the transverse velocity of the atom. Similar to \cite{1982ApJ...255..303L}, we decompose the velocity of the atom into two components: the velocity $v_1$ along the line of sight, and the velocity $v_2$ perpendicular to the line of sight. By applying a normalization factor, we introduce two transverse velocities: $u_1 = \sqrt{m/2kT} v_1$ and $u_2 = \sqrt{m/2kT} v_2$ that rescale the Maxwellian distribution to unit width. We will redistribute $u_1$ and $u_2$ separately, then calculate them into $x$, as well as the $\Delta \nu$ we are interested in.

\begin{figure}
\centering
\includegraphics[scale=0.65]{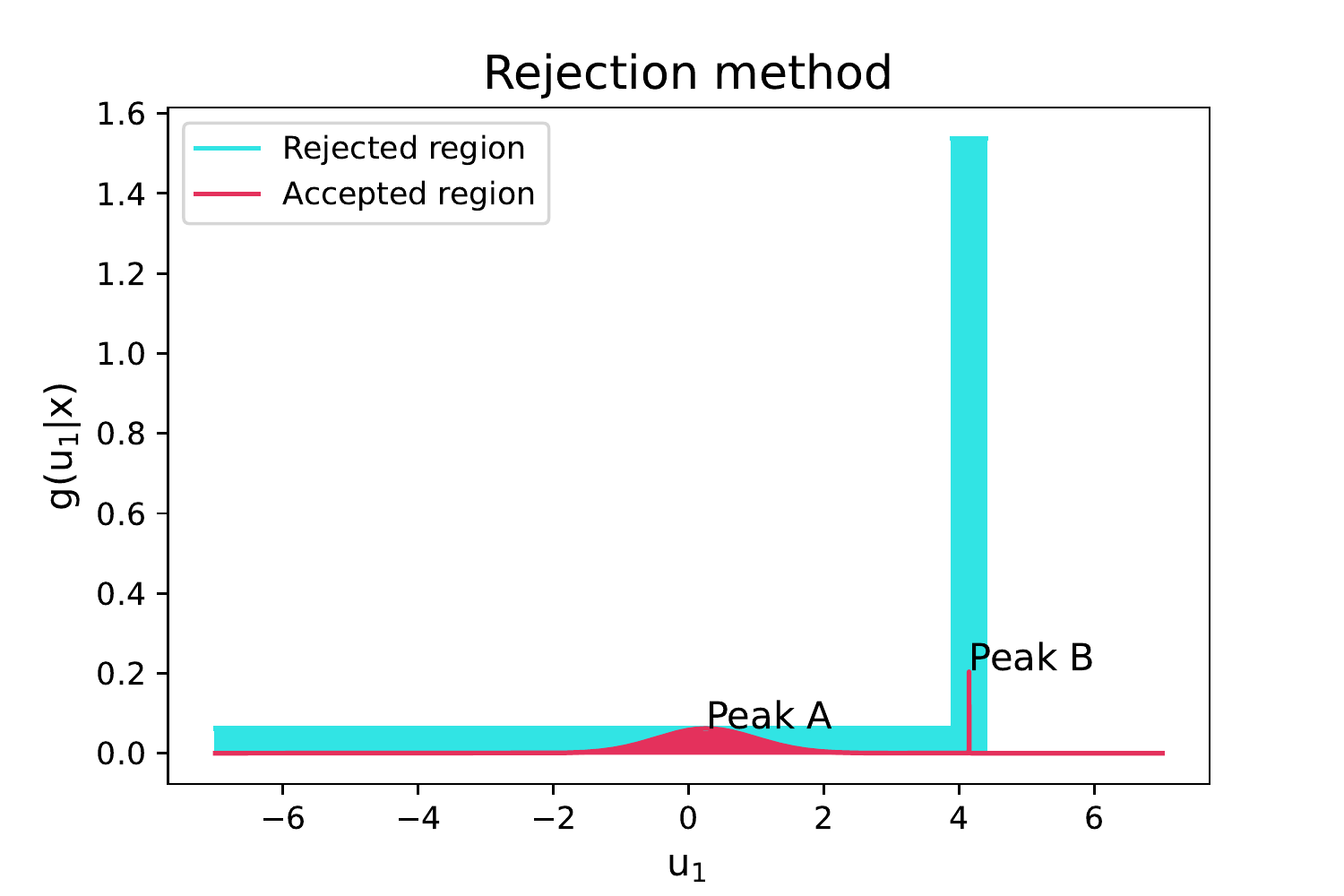}
\caption{\large Sample of rejected region and accepted region when $x = 4.14$. A random point is selected from the cyan rectangles (composed of 3 rectangles, of which 2 are large enough to be visible on the scale of the plot). It is accepted if it is below the $g(u_1|x)$ curve (maroon shaded region). \label{fig:RM}}
\centering
\end{figure}

First, we draw a random $u_1$. When $|x| \geqslant 6.5$, we choose two random numbers $r_1$ and $r_2$ between 0 and 1, then $u_1 = 1 / x + \sqrt{-\log{(r_1)}}\cos{(2\pi r_2)}$. We modify the derivation for $u_1$ when $|x|<6.5$. To use the rejection method, we want to build a region with 3 rectangles, where the horizontal endpoints are $u_{\rm 1a}=-7$, $ u_{\rm 1b} = x - 0.25$, $ u_{\rm 1c} = x + 0.25$, and $ u_{\rm 1d} = 7$. We ignore the cases that are outside this range because the probability of being outside is $\le 4.5 \times 10^{-19}$, which is negligible. For the height of each rectangle, we want to use the largest value of each section, and there are two cases. When the $|x|<2$, the largest values happen at the endpoints and we compare the two endpoint values $\exp(-u_1^2) / (u_1 - x)^2$ then choose the larger one. While when $|x|>2$, we need to set a midpoint $u_{\rm 1e} = (x - {\rm sign}_x \sqrt{x^2 - 4}) / 2$, and compare its $\exp(-u_{1e}^2) / (u_{1e} - x)^2$ to others and choose it if it any larger. After getting the width and height of each rectangle, we can compute the areas of them. Then we can compute the fraction of each area, which is also the probability of each region for a \Lya\ photon to be selected. Thus, in the rejection method, we can firstly use a random number to choose a rectangle, then use two random numbers to choose a point in that rectangle, and finally, test whether the point is under the curve $g(u_1|x)$ with corresponding $\rm u_1$. We use a {\tt while} loop to accept the $\rm u_1$ if it is under the curve and redo the whole process again if not. Figure~\ref{fig:RM} shows an example of the rejection method with rejection area and acceptance area when $x = 4.14$. Second, we draw a random $u_2$ from a Gaussian distribution, again using the polar transformation method.

\begin{figure}
\centering
\includegraphics[scale = 0.7]{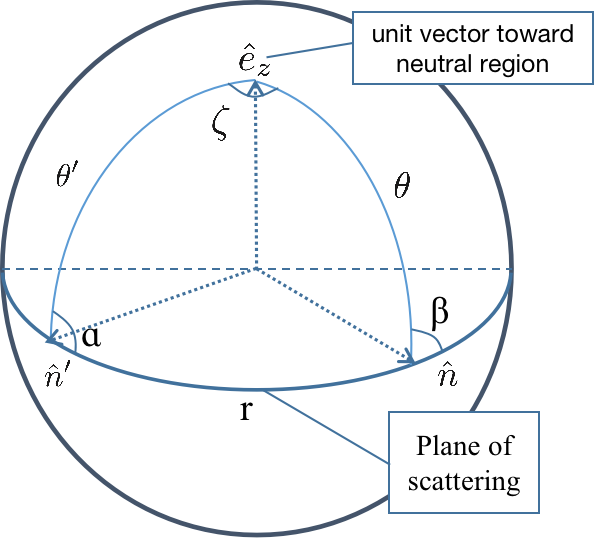}
\caption{\large Angles for calculating new direction} \label{fig:angles}
\centering
\end{figure}

Then we want to calculate the new direction and polarization for the \Lya\ photon together as these two properties are dependent. We describe and relate the initial and final directions of the scattering photon in spherical coordinates by six angles $\theta$, $\theta^{\prime}$, $\alpha$, $\beta$, $\gamma$ and $\zeta$ as shown in the Figure \ref{fig:angles}. The angle $\theta^{\prime}$ is the initial direction and $\theta$ is the final direction. For an isotropic distribution, the position angle $\alpha$ would be randomly chosen between 0 and $2\pi$ and the cosine of the scattering angle $\cos\gamma$ would be uniformly distributed between $-1$ and $+1$; we will use the isotropic distribution plus a rejection technique to draw from the full distribution. The final direction parameters $\theta'$, $\beta$, $\zeta$ is related to the initial direction $\theta$ and the two scattering geometric parameters $\alpha$ and $\gamma$ by spherical trigonometry:
\begin{eqnarray}
    \cos\theta &=& \cos\theta^{\prime} \cos\gamma + \sin\theta^{\prime} \sin\gamma \cos\alpha,
\nonumber \\
    \sin\theta \sin\beta &=& \sin\theta^{\prime} \sin\alpha,
\nonumber \\
\sin\theta \cos\beta &=& \sin\theta^{\prime} \cos\alpha \cos\gamma - \cos\theta^{\prime} \sin\gamma,
\nonumber \\
    \sin\theta \sin\zeta &=& \sin\alpha \sin\gamma, ~~{\rm and}
\nonumber \\
    \sin\theta \cos\zeta &=& \sin \theta^{\prime} \cos\gamma - \cos\theta^{\prime} \cos\alpha \sin\gamma,
\end{eqnarray}
and we can use rectangular-to-polar conversion to get $\theta$, $\beta$ and $\zeta$ individually. 

The new polarization could be derived from the direction information of outgoing photons. The fractional linear polarization $p$ is defined by
\begin{equation}
    p = \frac{I_{\rm NS} - I_{\rm EW}}{I_{\rm NS} + I_{\rm EW}},
\end{equation}
where $ I_{\rm NS}$ and $I_{\rm EW}$ represent the intensities that are measured in the North-South or East-West plane with the $z$-axis. The fractional linear polarization $p$ is between -1 and 1, where $-1$ indicates pure East-West polarization, $+1$ indicates pure North-South polarization, and 0 represents unpolarized state.

The probability distribution for scattering from direction $\boldsymbol{n^{\prime}}$ and polarization state $\rm q^{\prime}$ to direction $\boldsymbol{n}$ and polarization state $q$ is
\begin{equation}
    {\rm P}(\boldsymbol{n},q|\boldsymbol{n}',q') = \frac{1-E_1}{8\pi} + \frac{3}{8\pi}E_1 \cos^2\psi_{q,q'}
\label{eq:Pnn}
\end{equation}
where $\psi_{q,q'}$ is the angle between the initial and final polarization directions, and $ E_1$ indicates the scattering angular distribution \cite{1960ratr.book.....C}. Specifically, $ E_1 = 0$ refers to the pure isotropic scattering and $ E_1 = 1$ represents the pure dipole scattering. The value of $E_1$ depends on the angular momentum of the lower and upper levels. In this case, the admixture of isotropic versus dipole scattering depends on the frequency and the fine and hyperfine structure of the atom. The scattering process is of the form
\begin{equation}
{\rm H}(1{\rm s}_{1/2},F_{\rm i}) + \gamma \rightarrow {\rm H}(2{\rm p}_{j_{\rm e}},F_{\rm e}) \rightarrow {\rm H}(1{\rm s}_{1/2},F_{\rm f}) + \gamma,
\end{equation}
where $F_{\rm i}$, $F_{\rm e}$, and $F_{\rm f}$ are the total angular momenta of the initial, excited, and final states, respectively. There is an interference between the possible intermediate states ($j_{\rm e} = \frac12$ and $\frac32$, and $F_{\rm e} = j_{\rm e}\pm\frac12$) \cite{2006MNRAS.367..259H}.
We assume that the hydrogen atoms start with nearly random electron and nuclear spin, which is likely to be true in practice since all relevant temperatures are large compared to the hyperfine splitting ($k_{\rm B}\times 68$ mK). The problem simplifies in the case where the Voigt parameter $a\ll 1$ (natural width small compared to Doppler width), which is the case here. As can be seen from Figure~\ref{fig:RM}, scattering events can be either off resonance (Peak A) or on resonance (Peak B).

For events that are {\em off-resonance}, the frequency denominators $1/(E_{\rm i} + h\nu - E_{\rm e})$ in the scattering amplitude are essentially the same for all intermediate states. In this case, the electron and nuclear spin degrees of freedom are spectators, so the angular distribution is appropriate for angular momentum $0\rightarrow 1 \rightarrow 0$ scattering, i.e., $E_1=1$.
For events that are {\em on-resonance}, we find the averaged angular distribution parameter over the set of resonances, $\int \phi(\nu) E_1(\nu)\,d\nu$, and average over initial states with 1:3 weighting of $F_{\rm i}=0$ vs.\ $F_{\rm i}=1$ (\cite{2006MNRAS.367..259H}, Appendix B; note that there $\varpi_2 = \frac1{10}E_1$ is used instead). This yields $E_1=\frac13$ (in the limit that the 1s$_{1/2}(F=1)$--2p$_{3/2}(F=1)$ and 1s$_{1/2}(F=1)$--2p$_{3/2}(F=2)$ lines coincide; the offset is 23 MHz, which is less than the natural line width).
We implement this behavior in the code by setting
\begin{equation}
 E_1 = \begin{cases}
0 &  |x - u_1| \leq \sqrt3\, a \\ 1 & {\rm else}.
\end{cases}
\end{equation}
This way, in the off-resonance region, we have $E_1=1$, but in the on-resonance region we have $E_1$ either equal to 0 or 1, with an average of $\frac13$ taken over the Lorentzian distribution.

We can now write the probability distribution for each of the two final polarization states (NS or EW), for a partially polarized incident photon (arbitrary $p'$):
\begin{equation}
    {\rm P}(\boldsymbol{n},{\rm NS}|\boldsymbol{n}',p') = \frac{1-E_1}{8\pi} + \frac{3}{16\pi}E_1[(1+p')\cos^2\psi_{\rm NS,NS}+(1-p')cos^2\psi_{\rm NS,EW}]
\label{equ:3.15}
\end{equation}
and
\begin{equation}
    {\rm P}(\boldsymbol{n},{\rm EW}|\boldsymbol{n}',p') = \frac{1-E_1}{8\pi} + \frac{3}{16\pi}E_1[(1+p')\cos^2\psi_{\rm EW,NS}+(1-p')\cos^2\psi_{\rm EW,EW}].
    \label{equ.3.14}
\end{equation}
The angles can be inferred from spherical trigonometry; the explicit expressions are
\begin{eqnarray}
    \cos\psi_{\rm NS,NS} &=& \sin\theta \sin\theta' + \cos\theta \cos\theta' \cos\zeta,
\nonumber \\
    \cos\psi_{\rm NS,EW} &=& -\cos\theta \sin\zeta,
\nonumber \\
    \cos\psi_{\rm EW,NS} &=& \cos\theta' \sin\zeta,
    ~~~{\rm and}
\nonumber \\
    \cos\psi_{\rm EW,EW} &=& \cos\zeta.
\end{eqnarray}
The total probability is
\begin{equation}
    {\rm P}(\boldsymbol{n}|\boldsymbol{n}',p') = {\rm P}(\boldsymbol{n},{\rm NS}|\boldsymbol{n}',p') + {\rm P}(\boldsymbol{n},{\rm EW}|\boldsymbol{n}',p').
\end{equation} 

As the probability ${\rm P}(\boldsymbol{n}|\boldsymbol{n}',p')$ has a maximum possible value of $3/(8\pi)$, we also use the rejection method here. We generate a random number between 0 to 1, and accept the event if the random number is smaller than $(8\pi/3) {\rm P}(\boldsymbol{n}|\boldsymbol{n}',p')$, and otherwise draw again. The final polarization for that photon should be given by the relative probabilities to scatter into the NS or EW state, then the final polarization $p$ is
\begin{equation}
    p = \frac{{\rm P}(\boldsymbol{n},{\rm NS}|\boldsymbol{n}',p') - {\rm P}(\boldsymbol{n},{\rm EW}|\boldsymbol{n}',p')}{{\rm P}(\boldsymbol{n},{\rm NS}|\boldsymbol{n}',p') + P(\boldsymbol{n},{\rm EW}|\boldsymbol{n}',p')}.
\end{equation}

The algorithm to get the redistributed direction and polarization is:
\begin{enumerate}
\item Generate two random angles $\alpha$ and $\gamma$, then calculate $\theta$, $\beta$ and $\zeta$;
\item calculate $E_1$;
\item calculate the angles $\psi_{\rm NS,NS}$, $\psi_{\rm NS,EW}$, $\psi_{\rm EW,NS}$, $\psi_{\rm EW,EW}$, and then the probability densities ${\rm P}(\boldsymbol{n},{\rm NS}|\boldsymbol{n}',p')$ and ${\rm P}(\boldsymbol{n},{\rm EW}|\boldsymbol{n}',p')$;
\item calculate the total probability ${\rm P}(\boldsymbol{n}|\boldsymbol{n}',p')$;
\item generate a random number between 0 to 1, if the random number is smaller than $(8\pi/3) P(\boldsymbol{n}|\boldsymbol{n}',p')$, go next step, otherwise go back to step 1;
\item calculate $p$; and then
\item the final direction is $\theta$, and the final polarization is $p$.
\end{enumerate}

After getting scattering velocity and direction, we can calculate the new frequency offset (in Doppler units) $x$, which is
\begin{equation}
x_{\rm new} = x_{\rm i} + u_1\cos\gamma + u_2\sin\gamma - u_1.
\end{equation}
Multiplying by the local Doppler width, we get the new frequency offset $\Delta\nu = \sqrt2\,\sigma x$.

\subsection{Scoring of Monte Carlo}

After a \Lya\ photon is scattered and escaped, its final status will be recorded. Then we want to fit its direction probability density function with polynomial of direction $\mu$. We fit a Legendre polynomial expansion,
\begin{equation}
{\rm P}(\mu) = \sum_{j=0}^D b_j P_j(\mu),
\end{equation}
since the coefficients are more stable than for a regular polynomial expansion (using $\{\mu^j\}_{j=0}^D$ as a basis), and since the Legendre polynomial basis is natural when we go to a distribution of ionization front orientations (as we will in Paper II).
Here $P_j(\mu)$ is the Legendre polynomial of order $j$, $D$ is the order of the polynomial fit (our fiducial fit uses $D=5$), and there are $D+1$ coefficients $b_j$. The coefficients can be estimated as
\begin{equation}
b_j = \frac{2j+1}{2N} \sum_{i=1}^N P_j(\mu_i),
\label{eq:find-bj}
\end{equation}
where $i$ is summed over the $N$ photons propagated in the Monte Carlo, and $\mu_i$ is the direction cosine at which the $i$th photon emerges.\footnote{Since the Legendre polynomials are orthogonal, the linear algebra solution for the $b_j$ leads to a diagonal matrix. For purposes of generality, the implementation in our code does the matrix inversion.}

We also want to do the same thing for the mean polarization of corresponding $\mu$. To do that, we fit a model expansion for the linear polarization $\langle Q/I\rangle {\rm P}(\mu)$:
\begin{equation}
\left\langle \frac QI\right\rangle {\rm P}(\mu) =
\sum_{j=2}^D c_j P_j^2(\mu),
    \label{equ.3.23}
\end{equation}
where $P_j^2(\mu)$ is the associated Legendre function. The spin 2 basis functions are appropriate since linear polarization is a spin 2 field (recall that $_2Y_{j0}(\theta,\phi)$ is proportional to $P_j^2(\cos\theta)$ \cite{1967JMP.....8.2155G}). In particular, symmetry ensures that there is no linear polarization viewed from the $+z$ or $-z$ axis, i.e., $\langle Q/I\rangle {\rm P}(\mu)\rightarrow 0$ at $\mu=\pm 1$, and the use of the spin 2 basis functions enforces this. The coefficients $c_j$ can be estimated by an equation analogous to Eq.~(\ref{eq:find-bj}):
\begin{equation}
c_j = \frac{2j+1}{2 (j-1)j(j+1)(j+2)N} \sum_{i=1}^N p_i P_j^2(\mu_i).
\end{equation}
As the smallest $j$ here is equal to 2, we chose the highest degree $D = 6$.

\begin{figure}
\centering
\includegraphics[scale=0.7]{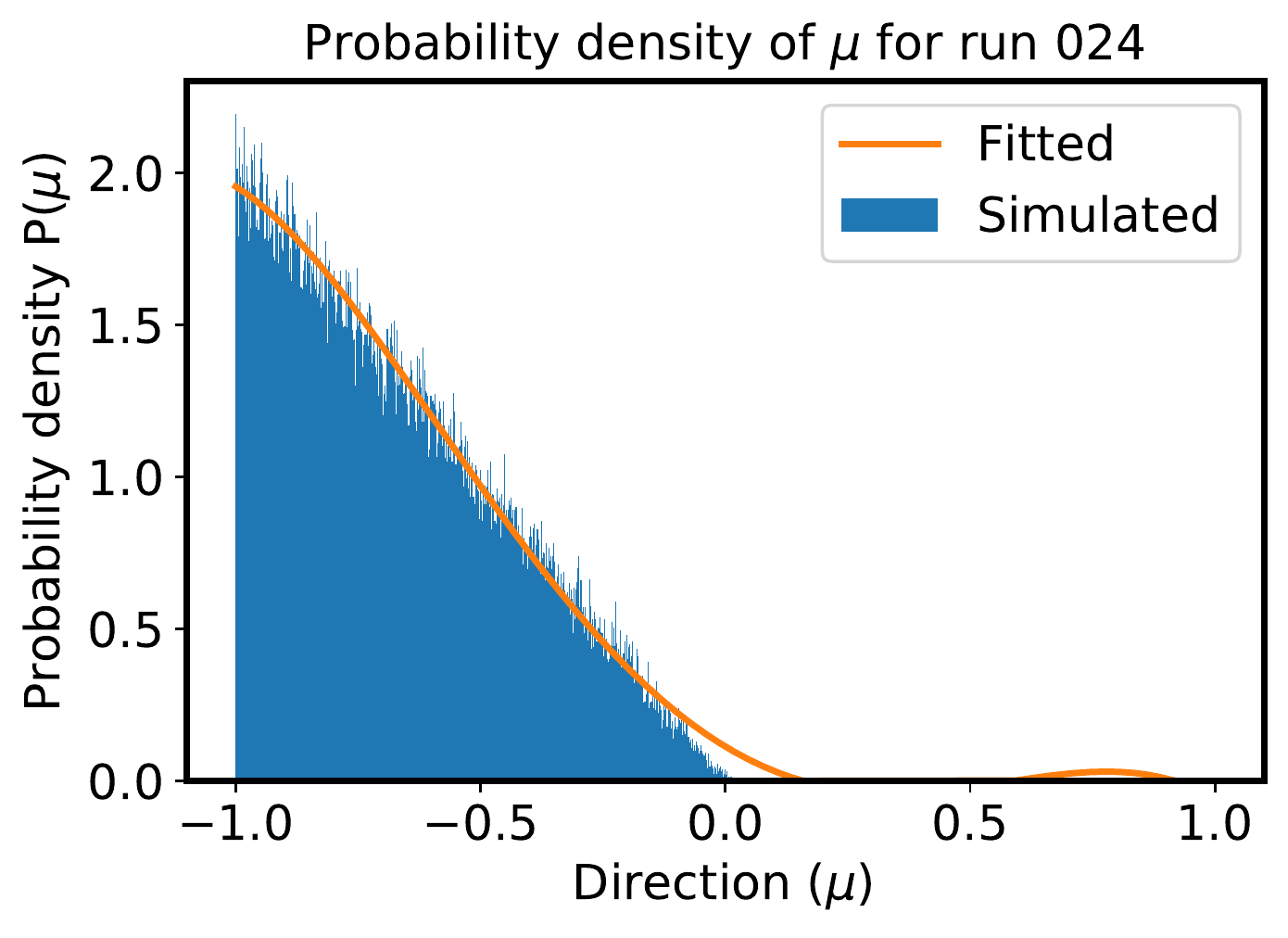}
\hfill
\includegraphics[scale=0.7]{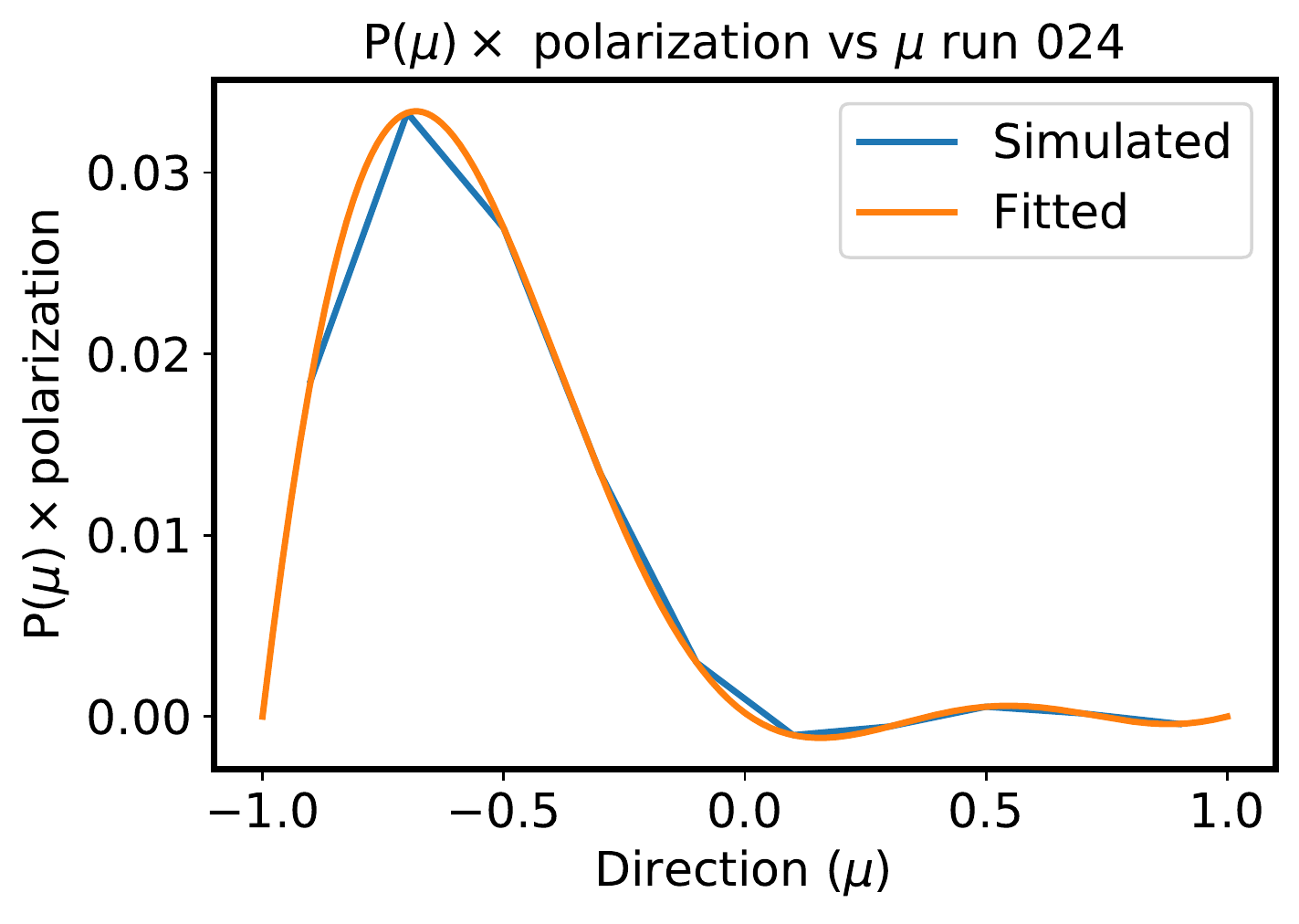}
\caption{\large Direction distribution (left) and mean polarization of each $\mu$ bin (right) of \Lya\ photons. The orange curve corresponds to the numerical fit we use in the following calculations, and the blue corresponds to the output of the datasets. \label{fig:prob density}}
\centering
\end{figure}

Figure $\ref{fig:prob density}$ shows an example of these polynomial fits.

\section{Relation to the Observer}

Given the probability distribution of photons emitted per unit area in some direction on the front, we can find the intensity of the front as given by 
\begin{equation}
    I_{\nu , \rm front} =\frac{hn \nu_{\rm front}{\rm P}(\mu)\delta(\nu_{\rm front} - \nu_{{\rm Ly}\alpha})}{2\pi |\cos\theta|} ,
    \label{equ.4.1}
\end{equation}
where $h$ is Planck's constant, $n$ is the rate of production of \Lya\ photons per unit area on the ionization front (units: photons cm$^{-2}$ s$^{-1}$), $z$ is the redshift of the front, ${\rm P}(\mu)$ is the probability density for the direction of emitted photons, and $\nu_{\rm front}$ is the frequency as observed when leaving the front, and $\theta$ is as defined above.
The rate $n$ is based on the temperature and the fraction of neutral hydrogen and free electrons:
\begin{equation}
    n = \sum_{i} q_{i} n_{{\rm HI},i} n_{e,i} s 
\end{equation}
where $i$ is a slab index, $s$ is the column density of hydrogen atoms in each slab given by $2.5 \times 10^{16}$ in $\rm cm^{-2}$, and $n_{\rm HI}$ and $n_{e}$ are the density of neutral hydrogen and free electrons in $\rm cm^{-3}$, respectively.
The variable $q$ is the rate coefficient in $\rm cm^3 \ s^{-1}$ for producing a \Lya\ photon from the 1s$\rightarrow$2p, 1s$\rightarrow$3s, and 1s$\rightarrow$3d electron-impact excitations. These rates are computed using
\comment{\begin{equation}
    \rm q = \frac{8.6287\times10^{-6} \gamma e^{-\Delta E/T_{e}}}{2 T_{e}^{1/2}},
\end{equation}
where $\Delta \rm E$ is the change in energy levels, $\rm T_{e}$ is the electron temperature and $\gamma$ is 
\begin{equation}
 \rm \gamma = a + bT_{e}+cT_{e}^{2}+dT_{e}^{4}   
\end{equation}
with coefficients given by }
Table 3 of Aggarwal \cite{1983MNRAS.202P..15A}.

We now apply Liouville's theorem to convert the specific intensity in the front frame to the observer frame. The conserved quantity is $I_\nu/\nu^3$ (see, e.g., \cite{1979rpa..book.....R}, \S4.9).
The intensity of the emitted photons an observer will see is related to the intensity as seen at the front by
$I_{\nu,\rm obs} = \nu_{\rm obs}^{3}/\nu_{\rm front}^{3}I_{\nu,\rm front} $.
Using $\nu_{\rm front} = (1+z)\nu_{\rm obs}$ and Eq.~(\ref{equ.4.1}), we find the intensity an observer would see is given by
\begin{equation}
    I_{\nu, \rm obs} = \frac{h\nu_{\rm obs}n {\rm P}(\mu) \delta ((1+z)\nu_{\rm obs} - \nu_{{\rm Ly}\alpha})}{2\pi |\cos\theta|(1+z)^{2}}
\end{equation}
which can be rewritten as 
\begin{equation}
    I_{\nu, \rm obs} = \frac{h\nu_{\rm obs}n {\rm P}(\mu) \delta (\nu_{\rm obs} - \nu_{{\rm Ly}\alpha}/(1+z))}{2\pi |\cos\theta|(1+z)^{3}}.
    \label{equ.4.6}
\end{equation}

We can similarly find the polarized intensity fraction by substituting the linear polarization in for the direction probability, i.e. P($\mu$) becomes $\langle Q/I\rangle{\rm P}(\mu)$ with $\langle Q/I\rangle$ defined by the polarization $p$ from Eq.~(\ref{equ.3.23}). If the position angle of the front in the sky is $\phi$, then we may further apply a rotation by angle $\phi$ to get the Stokes parameters in the observer frame:
\begin{eqnarray}
    Q_{\nu,\rm obs}&=& \frac{h\nu_{\rm obs}n p {\rm P}(\mu) \delta (\nu_{\rm obs} - \nu_{{\rm Ly}\alpha}/(1+z))}{2\pi |\cos\theta|(1+z)^{3}} \cos 2\phi
    ~~~{\rm and} \nonumber \\
    U_{\nu,\rm obs}&=& \frac{h\nu_{\rm obs}n p {\rm P}(\mu) \delta (\nu_{\rm obs} - \nu_{{\rm Ly}\alpha}/(1+z))}{2\pi |\cos\theta|(1+z)^{3}} \sin 2\phi.
    \label{equ.4.7}
\end{eqnarray}

\comment{Using the transformation matrix 
\[\begin{bmatrix}
\rm <Q/I>_{obs} \\
\rm <U/I>_{obs}
\end{bmatrix}=
\begin{bmatrix}
\rm cos(2\phi) && \rm sin(2\phi) \\
\rm -sin(2\phi) && \rm cos(2\phi) 
\end{bmatrix}
\begin{bmatrix}
\rm <Q/I>_{front} \\
\rm <U/I>_{front}
\end{bmatrix},
\]
we can find the relation of $\rm <Q/I>_{front}$ to $\rm <Q/I>_{obs}$ and $\rm <U/I>_{obs}$
By symmetry of our ionization front, we know $\rm <U/I>_{front}=0$, so this gives
\begin{equation}
\rm <Q/I>_{obs} = <Q/I>_{front} cos(2\phi)
\end{equation} and 
\begin{equation}
\rm <U/I>_{obs} = <Q/I>_{front} sin(2\phi)
\end{equation}}

\section{Results}

\begin{table}
\begin{center}
\begin{tabular}{| c || c | c | c|} 
 \hline
Index & Temperature (K) & Front speed (cm/s) & Hydrogen number density (cm$^{-3}$) \\ [0.5ex] 
 \hline\hline
 0 & $5\times 10^4$ & $5\times 10^7$ & $1.37\times 10^{-5}$  \\ 
 \hline
 1 & $6.25\times 10^4$ & $1\times 10^8$ & $5\times 10^{-5}$ \\
 \hline
 2 & $7.5\times 10^4$ & $5\times 10^8$ & $1.37\times 10^{-4}$ \\
 \hline
 3 & $8.75\times 10^4$ & $1\times 10^9$ & $5\times 10^{-4}$ \\
 \hline
 4 & $1\times 10^5$ & $5\times 10^9$ & $1.37\times 10^{-3}$ \\ 
 \hline\hline
\end{tabular}
\caption{\large Indexing scheme for each run based on temperature, front speed, and hydrogen number density. \label{table:runs}}
\end{center}
\end{table}

\begin{table}
\begin{center}
\begin{tabular}{||c c || c c|| c c||} 
 \hline
Run Number & Photons & Run Number & Photons & Run Number & Photons \\ [0.5ex] 
 \hline\hline
 014 & 193004 & 203 & 138066 & 333 & 121996 \\ 
 \hline
 023 & 149351 & 204 & 116466 & 334 & 148773 \\ 
 \hline
 024 & 108203 & 213 & 143613 & 343 & 101115\\ 
 \hline
 033 & 142671 & 214 & 119384 & 344 & 112404\\
 \hline
 034 & 115009 & 223 & 138846 & 402 & 148920 \\
 \hline
 043 & 154846 & 224 & 106993 & 403 & 101274\\
 \hline
 044 & 116031 & 233 & 127106 & 404 & 130431\\ 
 \hline
 103 & 177127 & 234 & 105027 & 412 & 152007 \\
 \hline
 104 & 147405 & 243 & 113118 & 413 & 103859\\
 \hline
 113 & 179377 & 244 & 125364 & 414 & 133432\\
 \hline
 114 & 146037 & 302 & 166469 & 422 & 178690\\
 \hline
 123 & 151398 & 303 & 115304 & 423 & 113517\\
 \hline
 124 & 114568 & 304 & 147525 & 424 & 138352\\
 \hline
 133 & 134830 & 312 & 172887 & 432 & 194819\\
 \hline
 134 & 108607 & 313 & 117248 & 433 & 115339\\
 \hline
 143 & 130951 & 314 & 100022 & 434 & 143541\\
 \hline
 144 & 144254 & 323 & 123828 & 443 & 144229\\
 \hline
 202 & 195440 & 324 & 148344 & 444 & 107601\\ [1ex]
 \hline
\end{tabular}
\caption{\large Number of photons in each associated run. Each run that is not listed above had exactly 100,000 photons associated with that simulation.}
\label{table: photon count}
\end{center}

\end{table}

\begin{figure}
    \centering
    \includegraphics[width=1\textwidth]{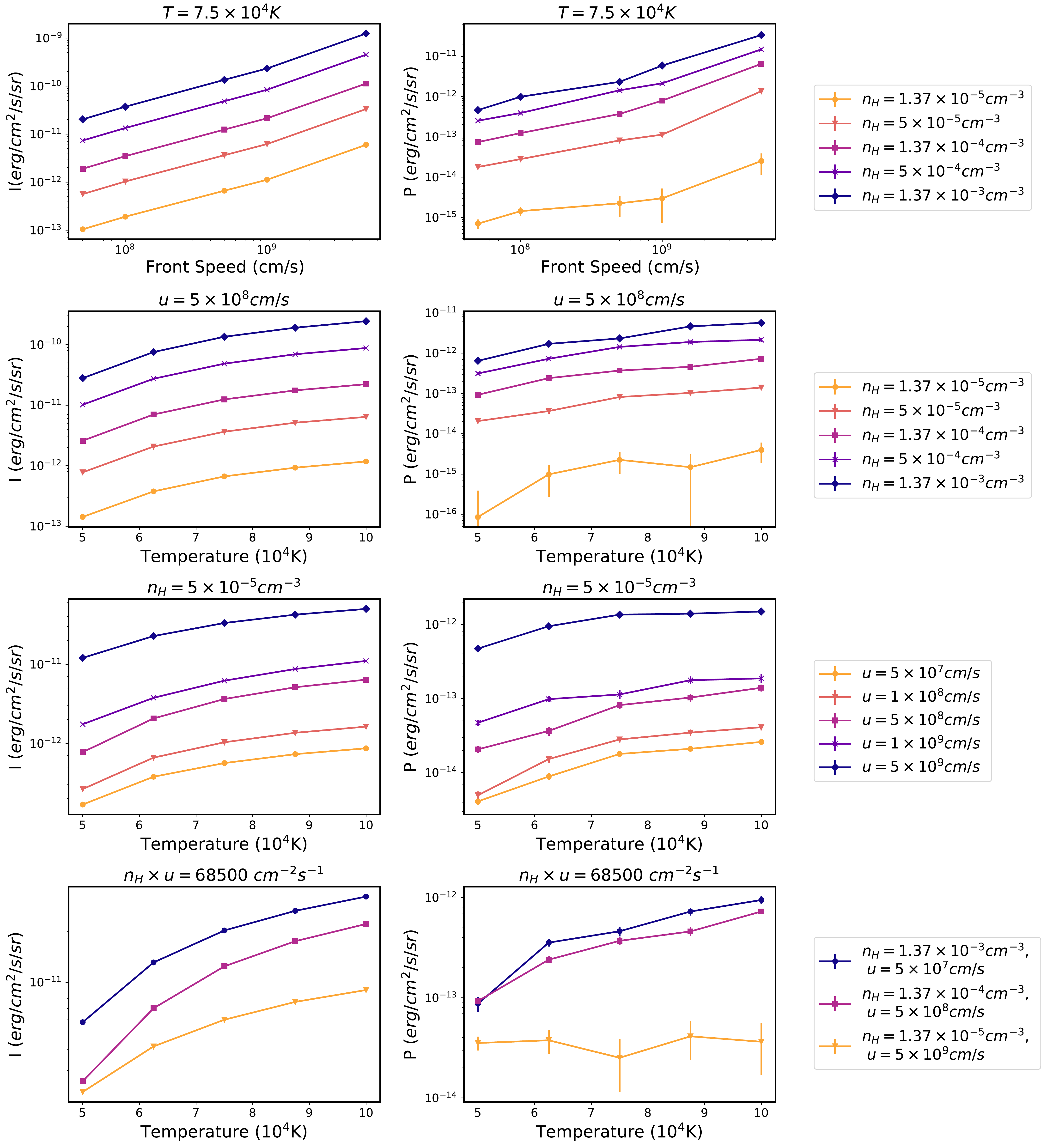}
    \caption{\large Trend lines in parameter space for intensity (left column) and polarized intensity (right column) all taken with $\mu=\cos(\frac{3}{4}\pi$). The first row shows 5 different $n_H$ values as a function of front speed at constant temperature. The second row shows 5 different $n_H$ values as a function of temperature at constant front speed. The third row shows 5 different front speed values as a function of temperature at constant $n_{\rm H}$. The last row is a diagonal slice in parameter space, with constant flux as a function of temperature.}
    \label{fig:results}
    
\end{figure}

\begin{figure}
    \centering
    \includegraphics[width=1\textwidth]{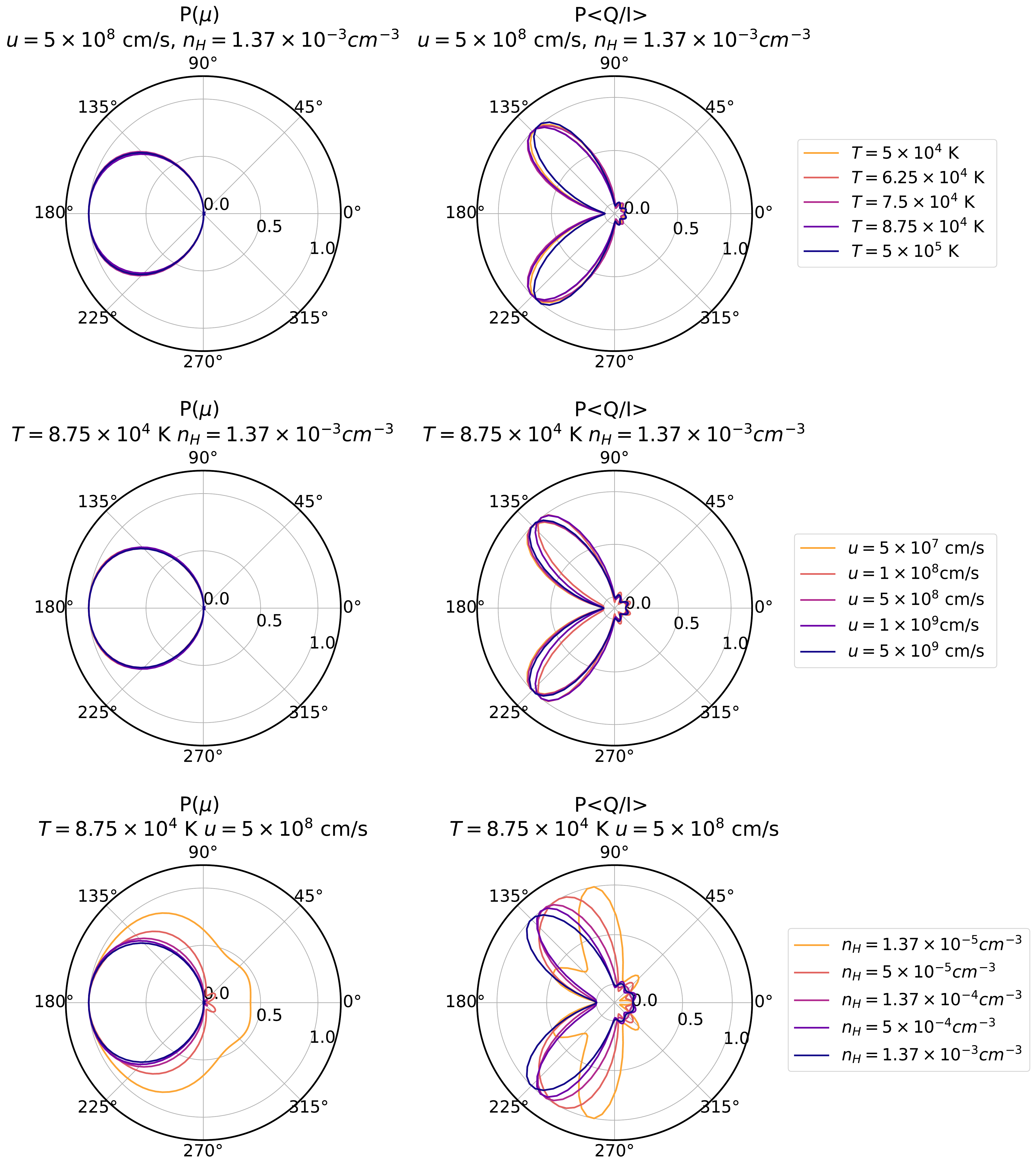}
    \caption{Renormalized polar plots as seen from the front to demonstrate directional dependence of intensities. The left (right) column demonstrates the (polarized) intensity as a function of direction. Each row shows a change in a single parameter; the first row shows changes in temperature, the second row shows changes in front speed, and the third shows changes in hydrogen density. We see the intensity and polarized intensity show a stronger dependence on hydrogen density than other parameters.}
    \label{fig:PolarPlots}
\end{figure}

\begin{figure}
    \centering
    \includegraphics[width=145mm]{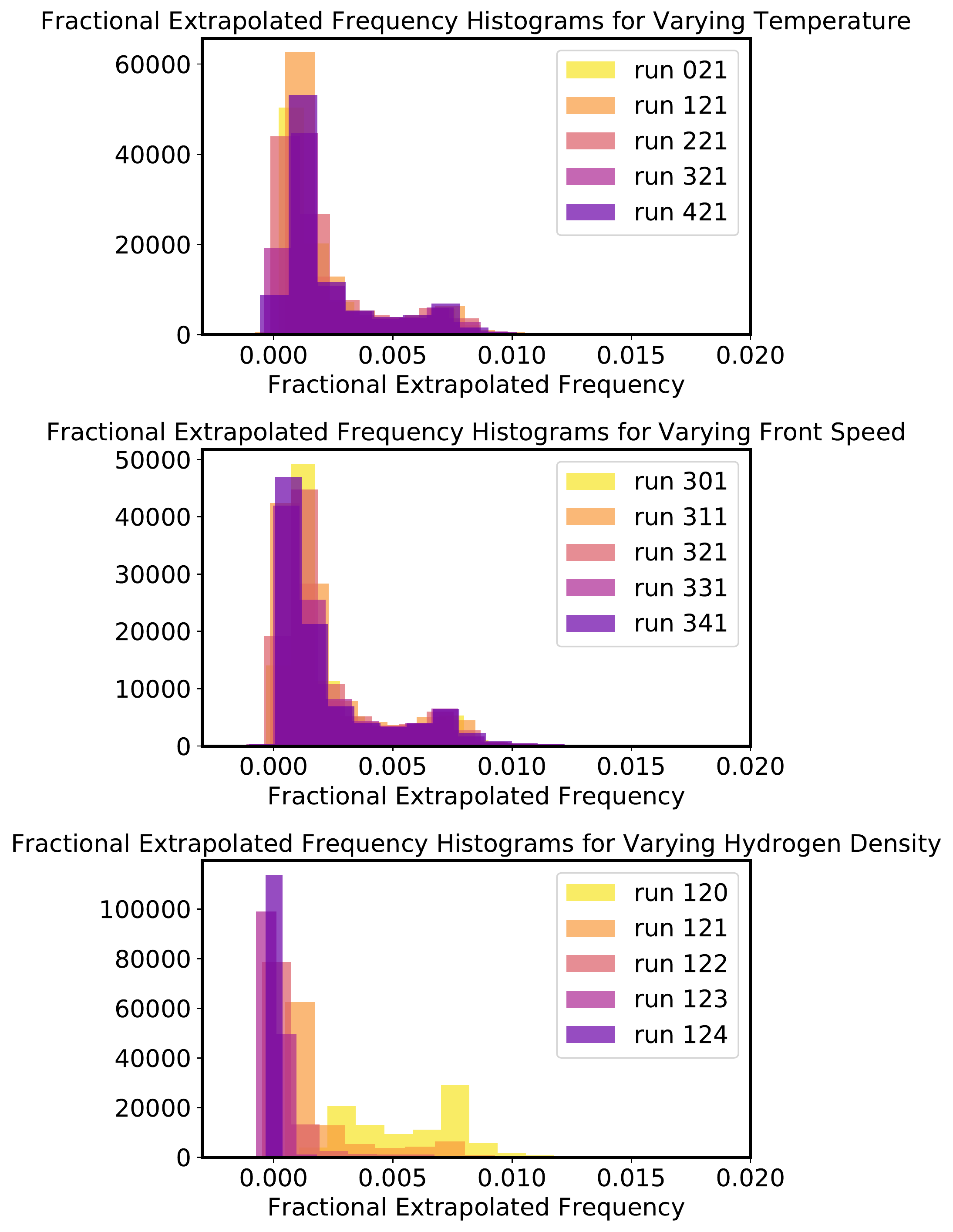}
    \caption{\large Extrapolated Frequency Histogram Examples. The top plot shows fractional extrapolated frequency for constant $U$ and $n_{\rm H}$, but varying temperature. The middle panel fixes the temperature and $n_{\rm H}$ and changes $U$. The last panel keeps temperature and $U$ the same and changes $n_{\rm H}$; this panel shows that the results are much more sensitive to $n_{\rm H}$ than the other parameters which show little deviation in each run.}
    \label{fig:three histograms}
\end{figure}

We show the results of our model across a grid of $5^3=125$ ionization front models, spanning a range of illumination blackbody temperature, ionization front speed, and hydrogen density.
The blackbody temperature is varied from $5\times 10^4$ to $10^5$ K, encompassing the range of potential cosmological ionization sources \cite{2001ApJ...552..464B}. The ionization front speeds range from $5\times10^{7}$ cm/s to $5\times10^{9}$ cm/s (for reference, a front at $5\times 10^8$ cm/s would advance by 4.9 cMpc per $\Delta z=1$, so this is a typical range for a cosmological ionization front) \cite{2019ApJ...874..154D}. The total hydrogen density is varied from $1.37\times10^{-5} \rm cm^{-1}$ to $1.37\times10^{-3} \rm cm^{-1}$, covering the range of 0.1 to 10 times mean density at $z=8$. Each run is indexed according to the scheme in Table~\ref{table:runs}: the first index accounts for temperature, the second corresponds to front speed, and the third indicates hydrogen density. Indices range from 0 for the lowest value to 4 for the highest value of each parameter.
The majority of runs were executed for 100,000 photons; the number of photons for runs that deviated from 100,000 photons can be found in Table~\ref{table: photon count}.

From these simulations, we calculated the intensity and polarized intensity for a given front, using Eqs.~(\ref{equ.4.6}) and (\ref{equ.4.7}) respectively. The intensity formula uses ${\rm P}(\mu)$ and the polarized intensity uses $p{\rm P}(\mu)$, which we have fit with the Legendre polynomial expansion and associated Legendre function expansion, respectively.

In Fig.~\ref{fig:results}, we display the values of intensity (left column) and polarized intensity (right column) in different slices of parameter space with a fixed value of $\mu = \cos(3\pi/4)$. The first row shows intensity of different hydrogen density curves as a function of front speed for constant source temperature  on a log-log plot. The second row displays intensity versus temperature for different hydrogen density values at a constant front speed. The third row fixes constant hydrogen density and varies the front speed. The fourth row shows intensity versus temperature for varying hydrogen density and front speed, but keeping the product of the two at a constant value of $6.85\times 10^4$ $\rm cm^{-2}\,s^{-1}$ to fix the ionizing flux (since the flux $F \sim n_{H}u$).

Generally we see that the polarized intensity is at least an order of magnitude smaller than the intensity. While near 100\%\ polarization is in principle possible from a collimated source scattering through a $90^\circ$ angle in the damping wings, the realistic combination of illumination and scattering geometries, and many scatterings occurring in the Doppler core of the line where the polarization is suppressed by a factor of $E_1=\frac13$, result in a much lower net polarization. We also see increasing trends in both total and polarizing intensities with higher temperatures, higher front speeds, and higher hydrogen densities. This makes sense since the illuminating flux is (aside from relativistic effects) proportional to the product of the gas density and front velocity, and a harder ionizing spectrum results in hotter gas and more Lyman-$\alpha$ cooling. There is a slight decrease in the average value of polarized intensity when varying front speed at the lowest hydrogen density, but as the strictly increasing trend is contained within one standard deviation, we do not take this to be a significant deviation.

For the constant product of hydrogen density and front speed, we see that more \Lya\ photons are produced by a slower but denser ionization region. As the density reduces and speed of the front increases, we see a decrease overall in the ionized \Lya\ photons produced. This could be understood by the fact that \Lya\ photons are created through collisional interactions, and with more hydrogen, there are more chances to scatter. There is also an increasing trend with temperature which is shown in all other plots as well. The polarized intensity demonstrates similar trends, but has the interesting result that the polarized intensity in the lowest density and highest speed seems to have a constant polarized intensity with respect to temperature. These fronts are being sampled at the same line of sight in this plot and tend to have an increasing maximum polarized intensity closer to perpendicular scattering angles, as seen in polar plots similar to figure \ref{fig:PolarPlots}. 

The error estimates in Fig.~\ref{fig:results} were calculated using the bootstrap resampling method for 1000 resamples of the Monte Carlo photons. We see the intensity is calculated to a smaller error than the polarized intensity, and in some of the polarized intensity cases, the error is on the same order of magnitude as the polarized intensity itself (in some cases, the result is consistent with zero polarization). We established 24 cases where the error of polarized intensity is greater than 20 percent of its value: 000, 010, 020, 030, 100, 110, 120, 130, 140, 200, 210, 220, 230, 240, 300, 310, 320, 330, 340, 400, 410, 420, 430, 440. In increasing the number of photons tested in these cases, we did not see any significant decrease in error.

In addition to mapping the intensity trends at a fixed $\mu$, we also looked at the directional dependence of intensity and polarized intensity. Three such plot pairs are given in figure \ref{fig:PolarPlots} where intensity and polarized intensities are normalized such that the maximum value is always one; this ensures we can compare distribution shape despite the intensity values being on different scales. An important feature of the polarized intensity plots is, because some polarization values are negative, there is a ``0 ring'' on the plot, and anything inside of that ring demonstrates a negative polarization (i.e., polarization in the East-West instead of North-South plane). While it is not clear whether these negative polarizations are numerical or physical in nature, the relative amplitude to the larger signal is so small that this will not impact future calculations. In the first row, we have fixed temperature and hydrogen density, and looked at the trend of increasing front speed. With higher front speed, we see the same distribution of intensity appearing. For the polarized intensity plot, we see a similar feature for all front speeds on the ionized side of some polarization perpendicular to the front, as well as some positive and negative polarized intensities but all very close to zero. On the other side of the front, we see near zero polarization perpendicular to the front. There are also symmetric bumps which change angle with increasing front speed, but all are close to 45 degrees from the front. In the next row, we plot the intensity distributions with constant front speed and incident blackbody temperature while allowing the hydrogen density to vary. For the intensity plot, we see a widening of the distribution from $180^\circ$ when we decrease the hydrogen density. This is especially prominent in the smallest hydrogen density case, where we see intensity of similar scales on both sides of the front. Since the scales for intensity are significantly smaller in the lowest hydrogen density case, we believe the difference in the shape of the trend can be accounted for with numerical effects from modeling the probability densities with curve fitting up to a finite order.

We also see very little polarized intensity on the neutral side of the front; again we see some additional variation in shape as we decrease the hydrogen density, which we determine to be from the smallest values of polarized intensity with largest error bars being associated with this hydrogen density. In the last row we look at the varying temperature plots, and again see that the distribution shapes for intensity and polarized intensity are not especially dependent on temperature. There is a bit of variation for the angle with maximum polarized intensity, but all occur at $\sim 135^\circ$, i.e., at an oblique angle as seen from the ionized side.

We also are interested in how narrow the ``$\delta$-function'' in frequency (Eq.~\ref{equ.4.1}) is as seen by a distant observer. To investigate this, we plotted the fractional extrapolated change in frequency for each photon given by 
\begin{equation}
     \frac{\Delta \nu_{\rm extrap}}{\nu_{{\rm Ly}\alpha}} = \frac{\Delta \nu_{\rm offset}}{\nu_{{\rm Ly}\alpha}} + \frac{H(z)(N_{\rm H, front} - N_{\rm H,exit})}{n_{\rm H} (c \mu - u)}.
\end{equation}
In this equation, $\Delta \nu_{\rm offset}$ is the photon's escape frequency subtracted by the initialized frequency, $H(z)$ is the Hubble constant at $z = 8$ in $\rm s^{-1}$, $ N_{\rm H,front}$ is the location where the front is 50\% ionized (as mesured by total hydrogen column density), $N_{\rm H,exit}$ is the photon's location at escape, $\rm n_{\rm HI}$ is the hydrogen density, $c$ is the speed of light, $\mu$ is the direction of the photon, and $\rm u$ is the front speed. 

Figure~\ref{fig:three histograms} displays the trends of the extrapolated frequency offset histograms while letting one of the parameters vary. The range of each plot is 2.5 times the standard deviation of the extrapolated frequency offset. 
For the cases of varying temperature or front speed, we see very little deviation in shape or spread of the distributions. By varying the hydrogen density, we see different patterns emerging. For the lowest hydrogen density, we see a very strong bimodal peak with one peak close located at 0.002852  and the stronger peak at 0.007605 on the bluer side, with the middle 80\% located in a range of 0.005788. As we view the next smallest hydrogen density, we see still a bimodal peak but now the bluer peak is suppressed in favor of the peak close to 0, located at 0.007392 and 0.001098 respectively, and the range of the histogram is given as 0.005967. As we again increase hydrogen density, we see the two peaks collapse into a single peak located at 0.0001073, and the range is 0.001143. This stronger-peak, smaller-spread distribution trend continues as we continue to increase the hydrogen density, eventually ending in a range of $5.144\times10^{-5}
$.

In each trial, there were 0--50 photons that had a fractional extrapolated frequency greater than the 2.5 standard deviation range. These photons have a value of $\mu\approx U/c$, which makes the second term of the extrapolated frequency equation become quite large. Physically, this is because if an ionization front is observed along the direction $\mu=U/c$, it is oriented in the ``radial direction'' (observer within the plane of the front) in light-cone coordinates. So if a photon appears at a transverse position that is a bit different from where the front is drawn, the observer assigns it to a radial position that is very far in front of or behind the position where it originated. These photons are not distinguishable from other points on the ionization front at a larger or smaller redshift, and do not contribute to the broadening of the front in redshift space. 

\section{Discussion}

With the model presented in this paper, we can calculate the total and polarized intensity of the Lyman-$\alpha$ emission from ionization front as a function of the basic physical inputs: the incident radiation spectrum; the gas density; the speed of the ionization front; and the viewing geometry of the observer. The model contains a detailed treatment of the physics of the ionization front, including (i) H and He ionization structure, including tracking the attenuation of each frequency bin in the incident spectrum; (ii) thermal evolution and collisional Lyman-$\alpha$ production rates for a multi-temperature plasma; (iii) Monte Carlo treatment of the photon propagation through the ionization front, including redshifting and scattering with frequency distribution; and (iv) tracking of the photon polarization, including the joint polarization-angular-frequency dependence of the scattering cross section. We have constructed a grid of models and explored the dependence of the intensity and polarized intensity on the model parameters.

Our remaining goal is to go from the emergent polarized intensities computed from these grids to the observable power spectrum of Lyman-$\alpha$ emission from the ionization fronts. This requires us to run a simulation of reionization in a cosmological volume, identify the ionization fronts, interpolate from our grid to compute the Lyman-$\alpha$ polarization in each cell, and finally run a power spectrum estimator on the simulated box. In Paper II, we will carry out this procedure using a {\sc 21cmFAST} \cite{2011MNRAS.411..955M, 2020JOSS....5.2582M} simulation, and assess the detectability of the Lyman-$\alpha$ polarization with plausible future experiments.

\section*{Acknowledgements}
We thank Tzu-Ching Chang for useful comments on the draft. During the preparation of this work, the authors were supported by NASA award 15-WFIRST15-0008, Simons Foundation award 60052667, and the David \& Lucile Packard Foundation.

This article used resources on the Pitzer Cluster at the Ohio Supercomputing Center\cite{OSC}.

\section*{Data Availability}

The code and data supporting this article may be made available on reasonable request to the corresponding author. 

\appendix

\section{Test cases for the Monte Carlo code}
\label{app:tests}

This appendix describes the suite of test cases that we have run for the Monte Carlo code.

\subsection{Photon generation}
\label{test:generate}

\begin{figure}
\centering
\subfloat[]{\includegraphics[width=75mm]{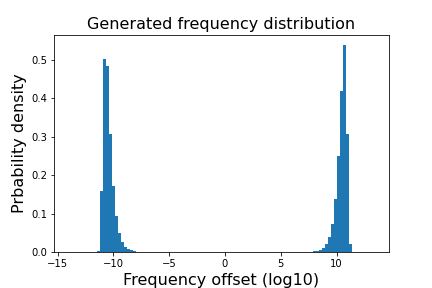}
\label{fig:vi}}
\subfloat[]{\includegraphics[width=75mm]{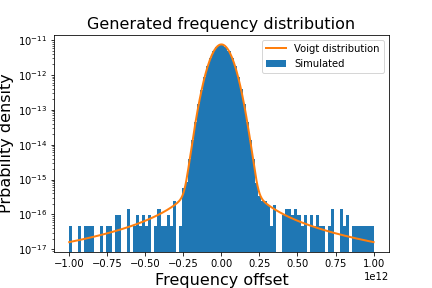}
\label{fig:ini}}
\caption{\large Distribution of generated frequency}
\label{fig:vi_ini}
\end{figure}

To begin, we test the photon generating process by exploring the frequency generator. Figure \ref{fig:vi_ini} shows the distribution of the generated frequency of photons. This test runs $10^6$ photons under the setting of background temperature of hydrogen atom $T_{\rm HI} = 5000$\ K. The Doppler width $\sigma$ in Eq.~(\ref{eq.G}) is about $5.29 \times 10^{10}$ Hz. Thus, the absolute value of generated $\Delta \nu$ is expected to be around $10^{10}$ Hz. The distribution in figure \ref{fig:vi} corresponds to the expectation. Further, after we change our base to linear horizontal axis and log vertical axis, the frequency distribution shown in figure \ref{fig:ini} matches the Voigt distribution, which also meets our expectation.

\subsection{Photon propagation}

\begin{figure}
\centering
\subfloat[]{
\includegraphics[width=74mm]{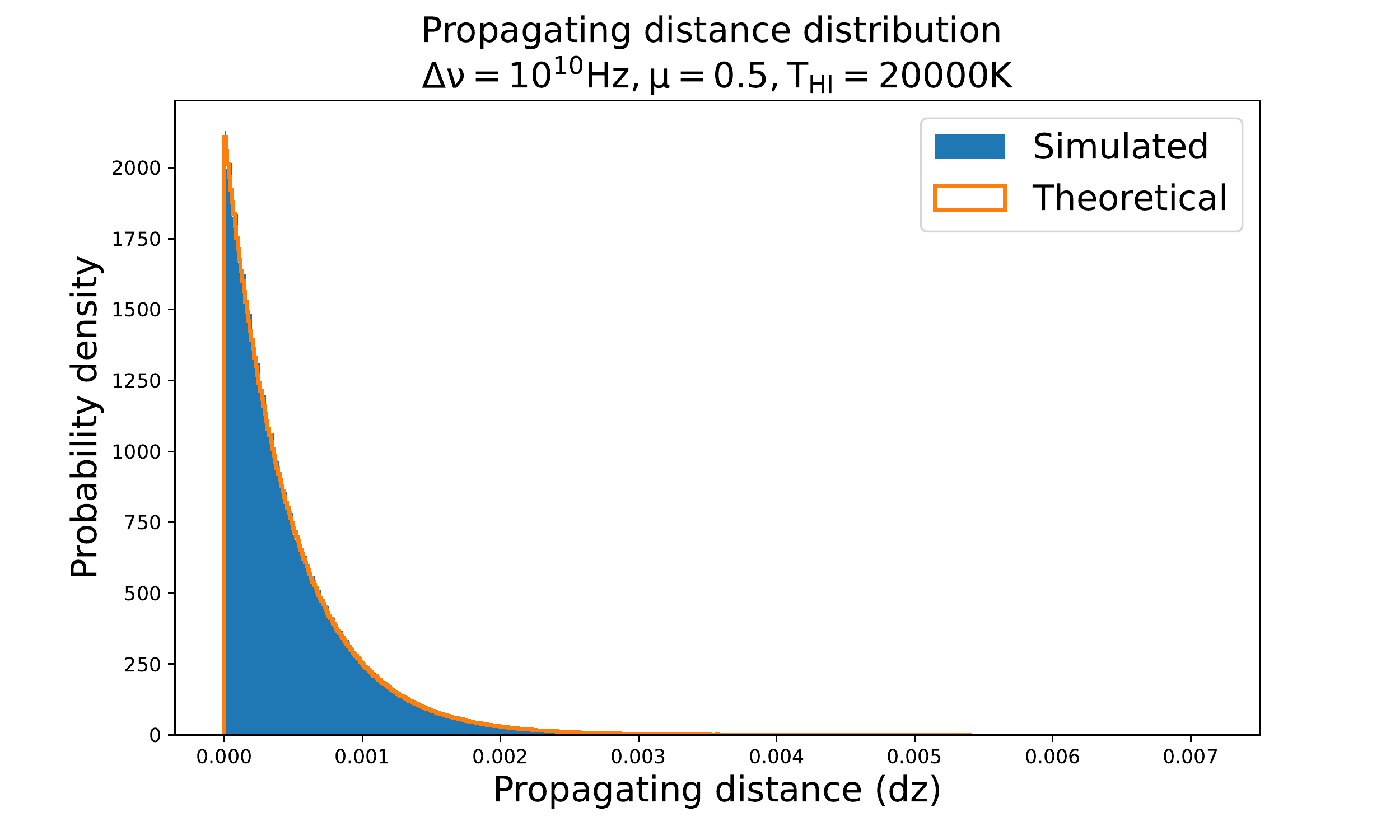}
\label{fig:pro_pos}}
\subfloat[]{
\includegraphics[width=74mm]{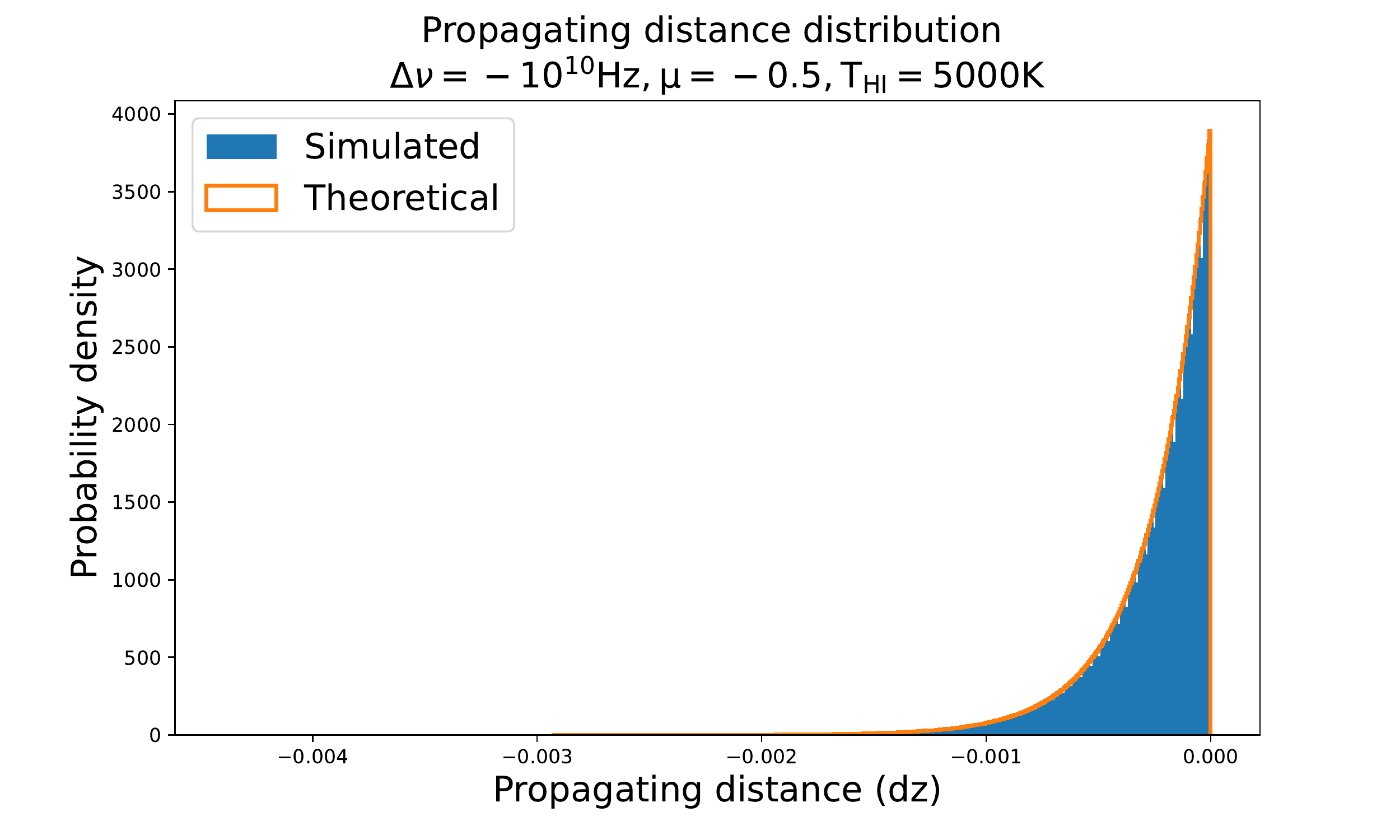}
\label{fig:pro_neg}}
\caption{ Distribution of propagation distances (projected onto the $z$-axis): (a) toward the neutral side and (b) toward the ionized side. In both cases, the distribution follows the expected exponential distribution.
\label{fig:propagate}}
\end{figure}

Here we test the propagating process. In order to do this test more precisely, we test both propagation toward the neutral side ($\mu>0$) and toward the ionized side ($\mu<0$). Figure \ref{fig:propagate} shows two cases: (a) \dv\ = $10^{10}$ Hz, $\rm cos\theta = 0.5$ and $T_{\rm HI} = 20000 \rm K$; and (b) \dv\ = $-10^{10}$ Hz, $\rm cos \theta = -0.5$ and $T_{\rm HI} = 5000 \rm K$. Each test runs $10^7$ photons. The orange line shows the theoretical distribution calculated corresponding to Section \ref{sec:PP}, and it is consistent with the simulation results.

\subsection{Scattering frequency}

To test the scattering process, we need to test the scattering frequency, scattering direction, and scattering polarization results. We start with the scattering frequency tests. Note that the scattering frequency test generally uses the normalized frequency offset $x$ and projection of the atom velocity along the incoming photon direction $u_1$, rather than \dv.

\begin{figure}
\centering
\subfloat[]{
\includegraphics[width=100mm]{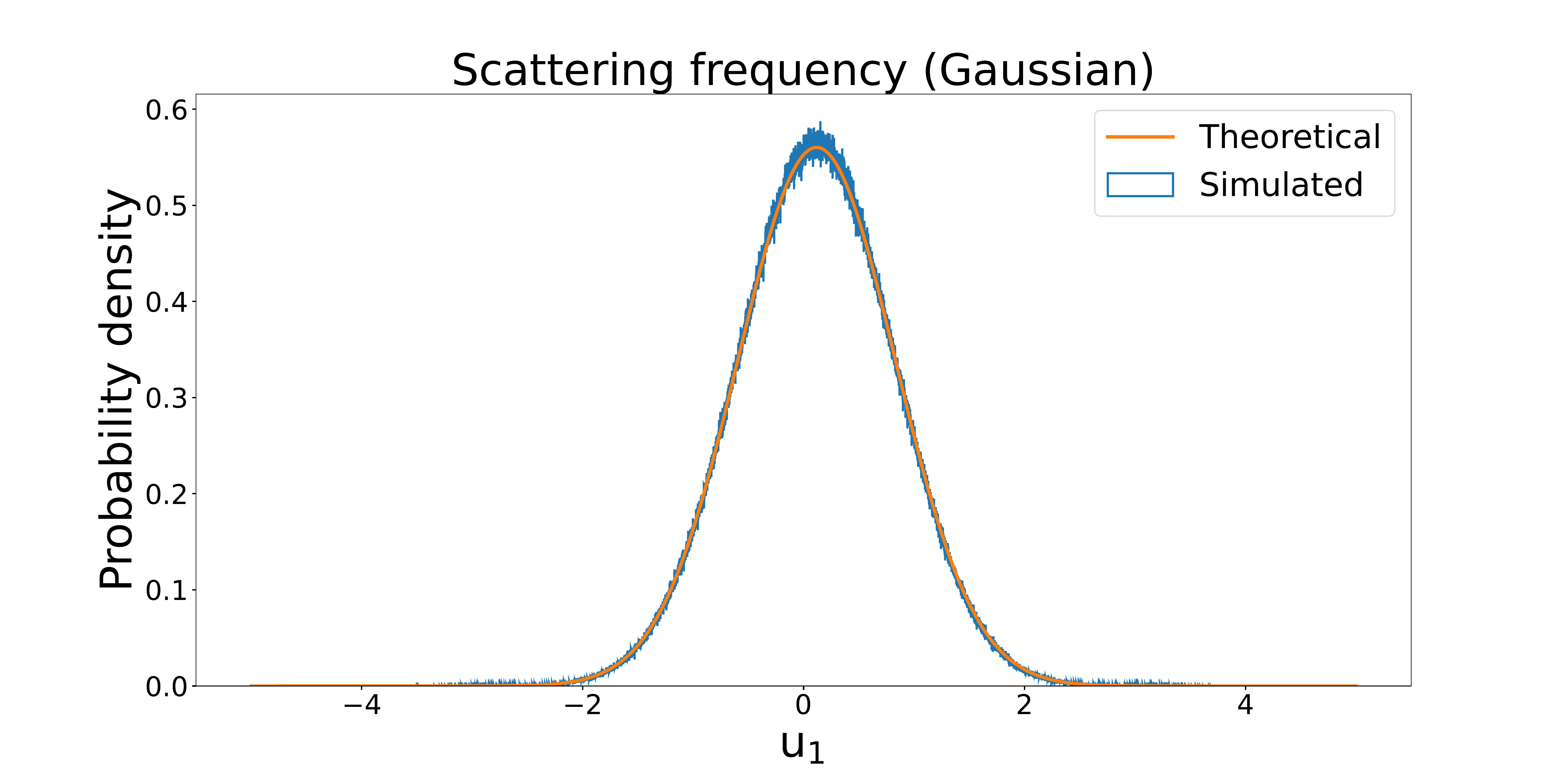}
\label{fig:SG}}
\hspace{0mm}
\subfloat[]{
\includegraphics[width=75mm]{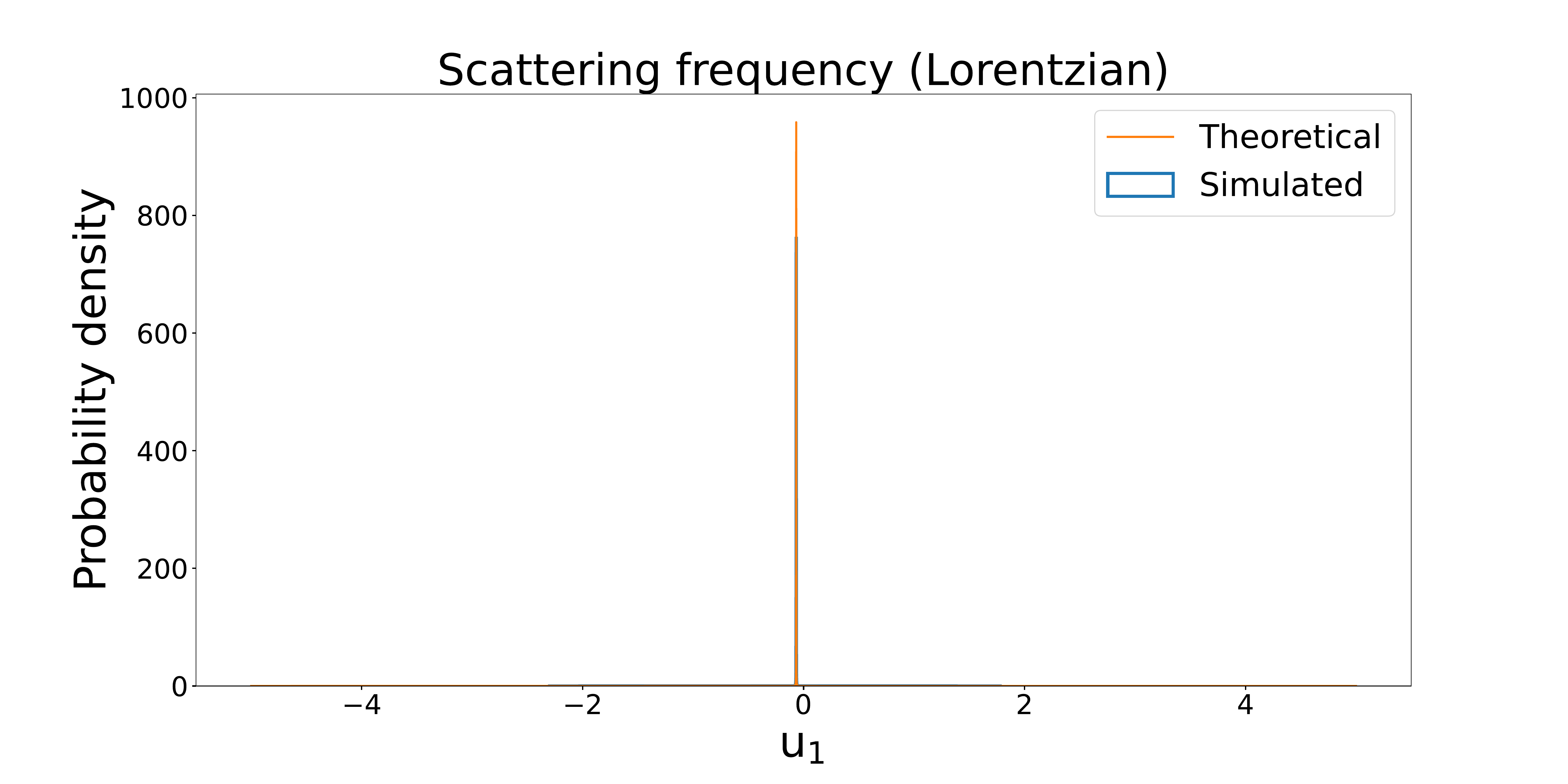}
\includegraphics[width=75mm]{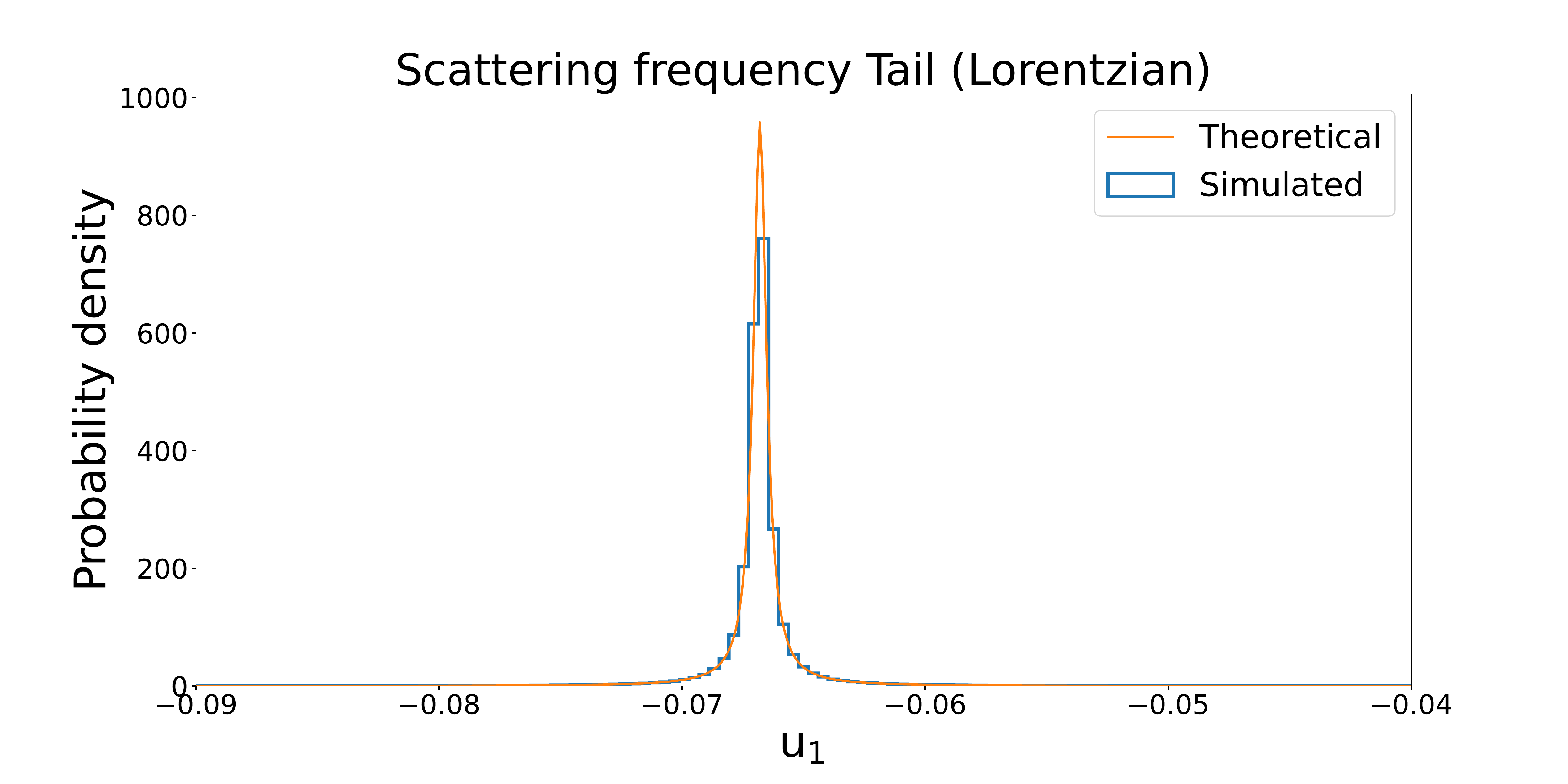}
\label{fig:SL}}
\hspace{0mm}
\subfloat[]{
\includegraphics[width=75mm]{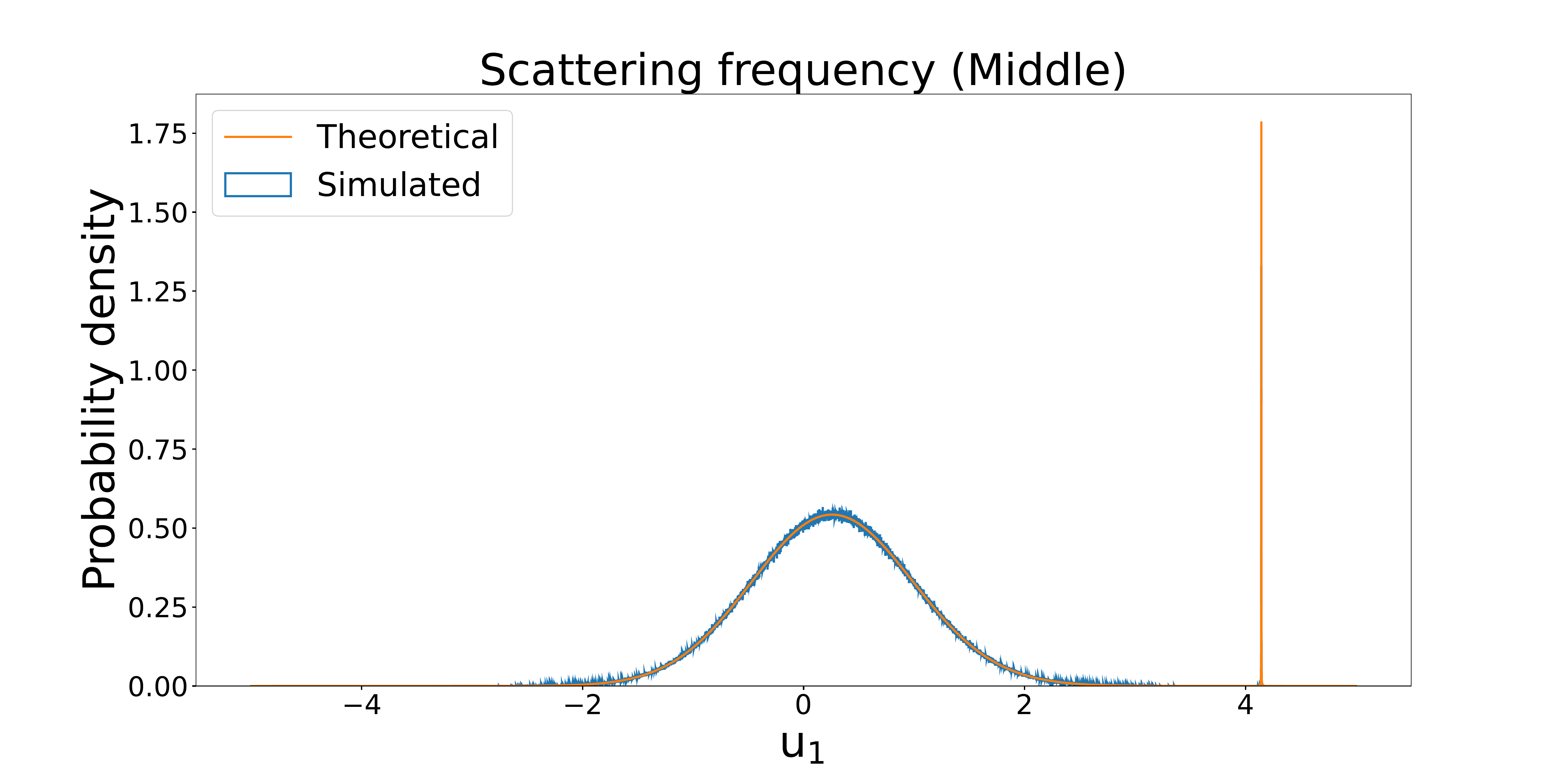}
\includegraphics[width=75mm]{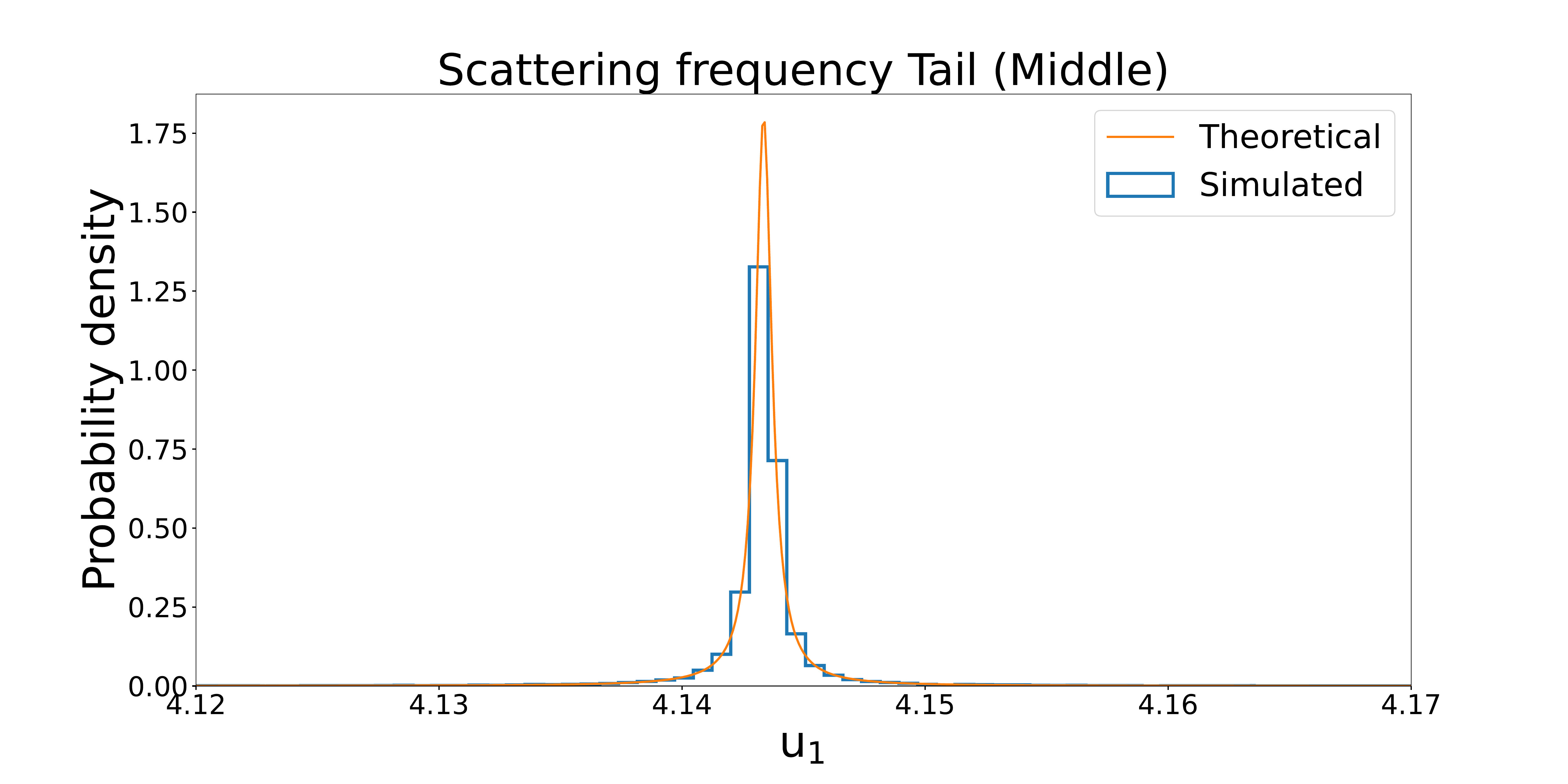}
\label{fig:SM}}
\caption{{\em Blue curves}: The simulated probability density of atomic velocity $u_1$ for (a) the Gaussian limit (large $|x|$); (b) the Lorentzian limit (small $|x|$); and (c) an intermediate bimodal case. {\em Orange curves}: The analytic prediction for the probability distribution. For the Lorentzian and intermediate cases, the right part of the figure shows a zoom-in of the $u_1\approx x$ region. 
\label{fig:scatv}}
\end{figure}

To test the selection of scattering atom velocity, we test the Gaussian (larger $|x|$) conditions, Lorentzian (smaller $|x|$) conditions, and the combination Gaussian and Lorentzian (middle $|x|$) limiting cases for ${\rm P}(u_1|x)$. Figure~\ref{fig:SG} shows the Gaussian limit with \dv\ = $-10^{12}$ Hz and $T_{\rm HI} = 12000$\,K; here $x=8.627$, the incoming photon is in the red damping wing of the Lyman-$\alpha$ line, and ${\rm P}(u_1|x)$ is a small perturbation on the Maxwellian distribution $\propto e^{-u_1^2}$. Figure \ref{fig:SL} shows the Lorentzian limit with \dv\ = $10^{10}$ Hz and $T_{\rm HI} = $20000\,K; here $x=-0.067$, and scattering occurs almost entirely off of hydrogen atoms whose velocities are such that the photon frequency in the atom frame is on the Lyman-$\alpha$ resonance. The result is a Lorentzian distribution of width $a$ centered at $u_1=x$. Figure \ref{fig:SM} shows an intermediate case with \dv\ = $-5 \times 10^{11}$ Hz and $T_{\rm HI} = $13000\,K; here $x=4.143$ and we see a bimodal distribution, with one broad peak neat $u_1\approx 0\pm 1$ (where most of the \HI\ atoms are located) and a narrow peak at $u_1\approx x\pm a$ (where the photon is resonant in the atom frame). 
In this test, about 79.73\% of photon is in the broad peak, 0.04\% of photon is in the narrow peak, and 20.23\% is distributed between $u_1\approx$ -3.358 to $u_1\approx$ 4.334. We set both the initial direction and initial polarization to 0 for all three cases, and run $10^7$ photons for each case. The orange lines are the theoretical distribution according to Lee \cite{1977ApJ...218..857L} (notice that the Voigt function here should be divided by $\sqrt{\pi}$ to be normalized to 1), and they are consistent with the simulation results.

\subsection{Scattering direction and polarization}

\begin{figure}
\centering
\includegraphics[width=75mm]{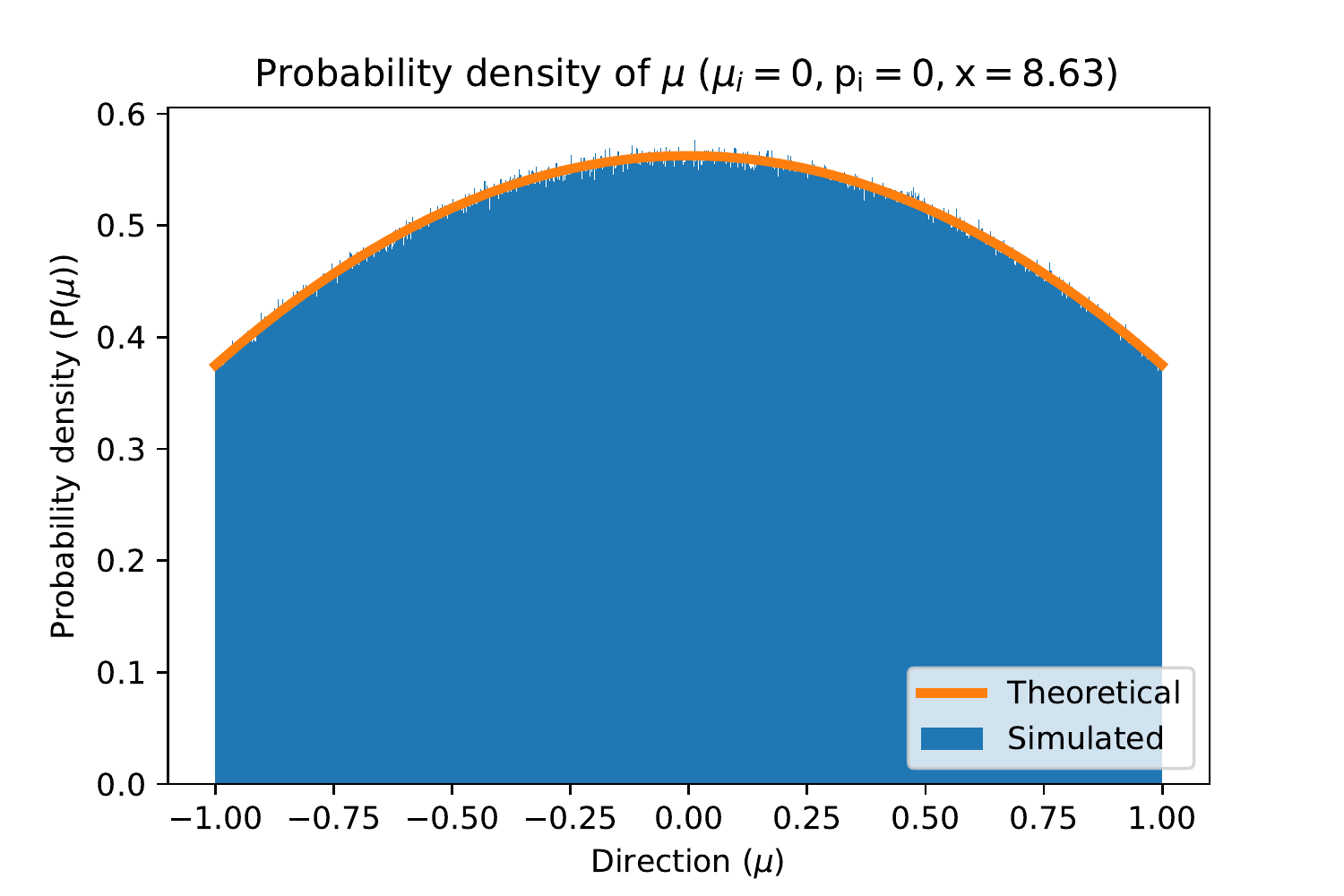}
\includegraphics[width=75mm]{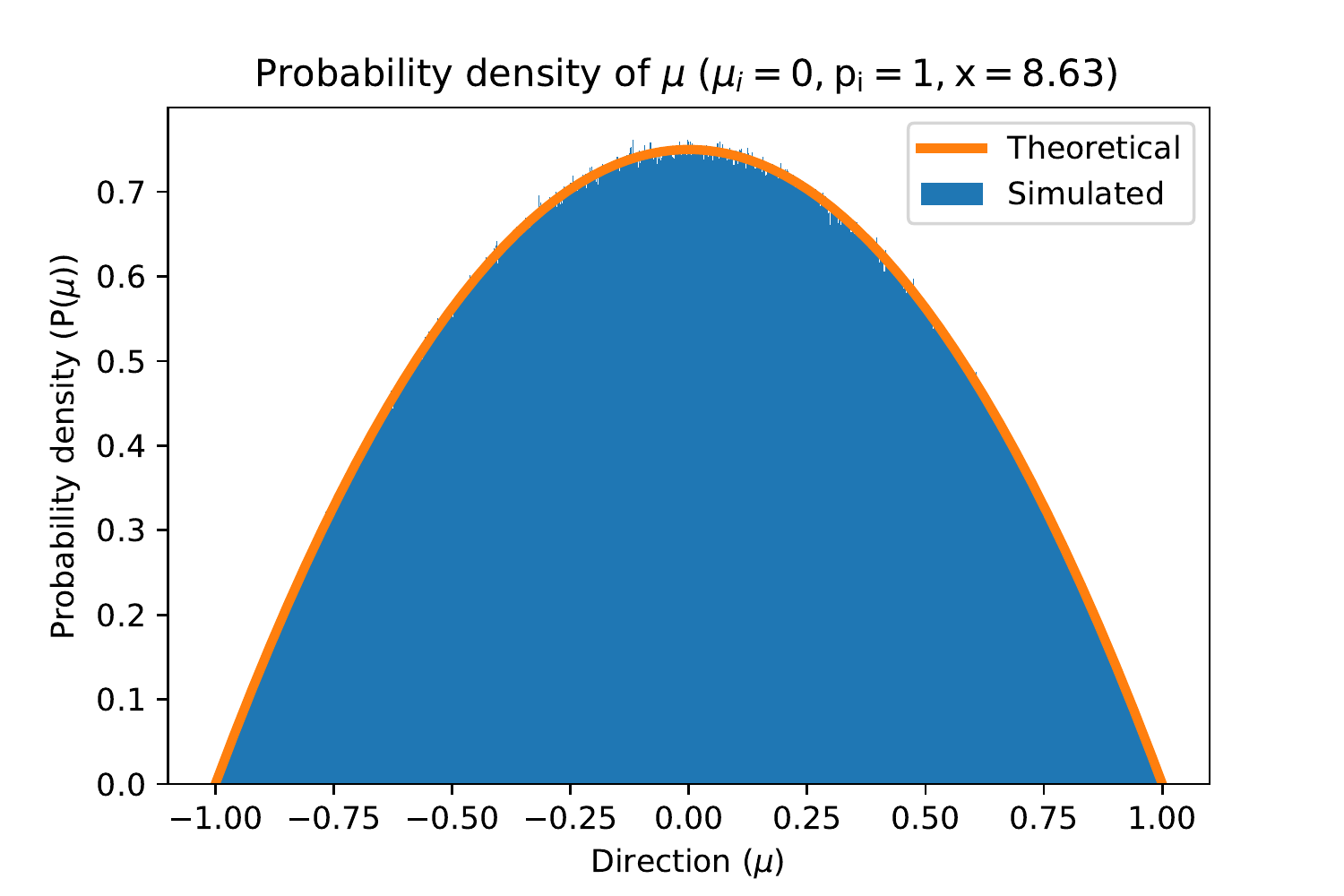}
\includegraphics[width=75mm]{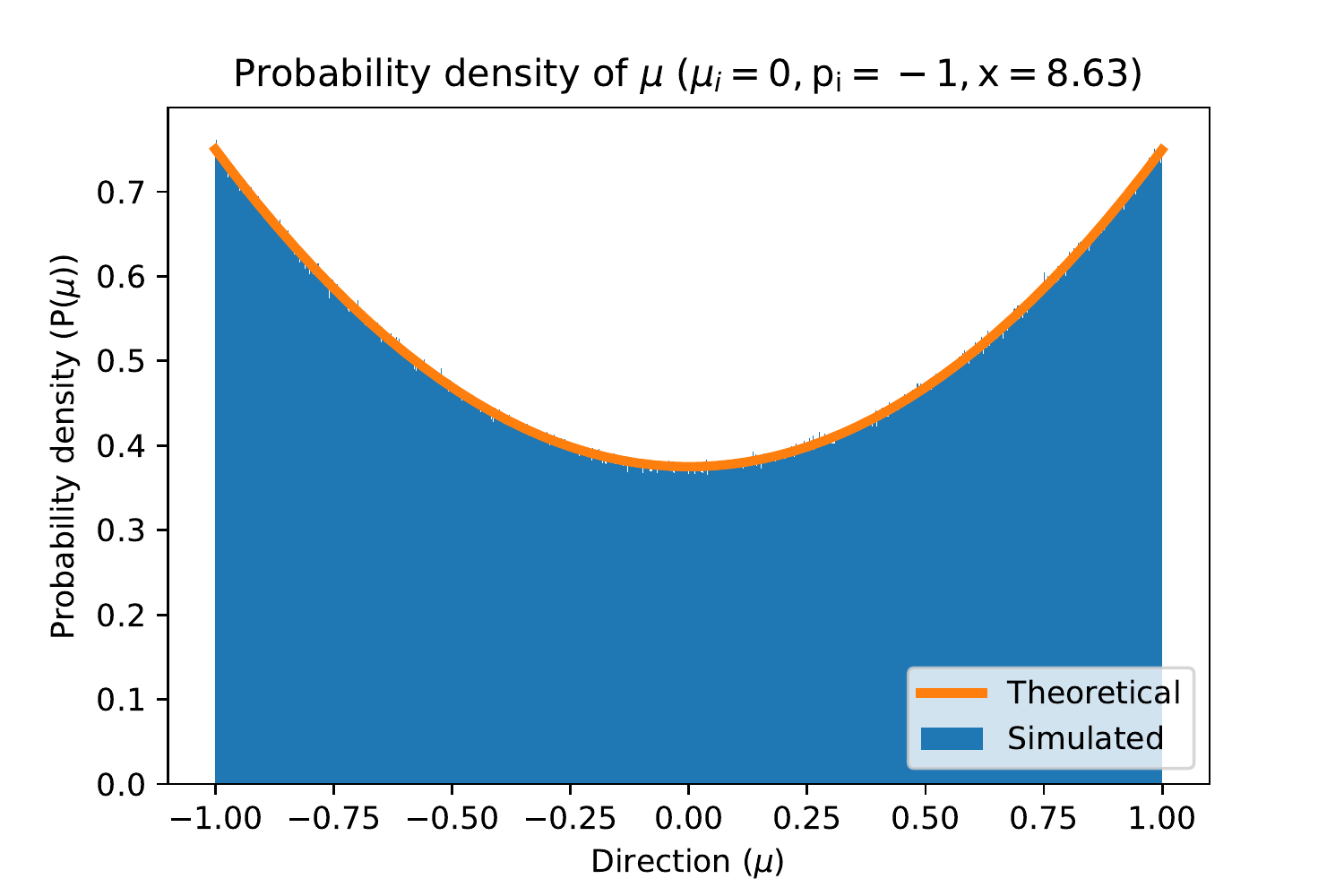}
\includegraphics[width=75mm]{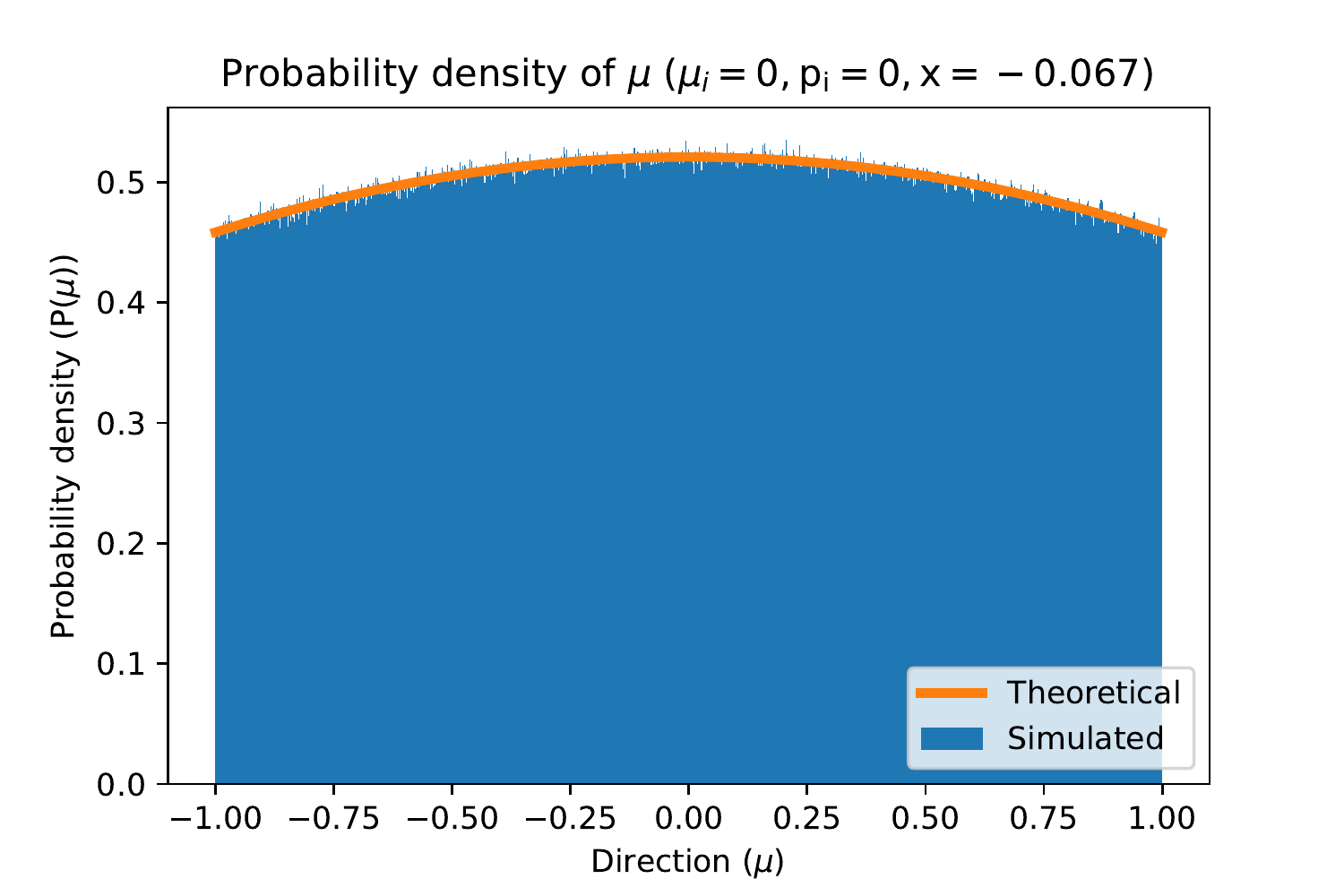}
\includegraphics[width=75mm]{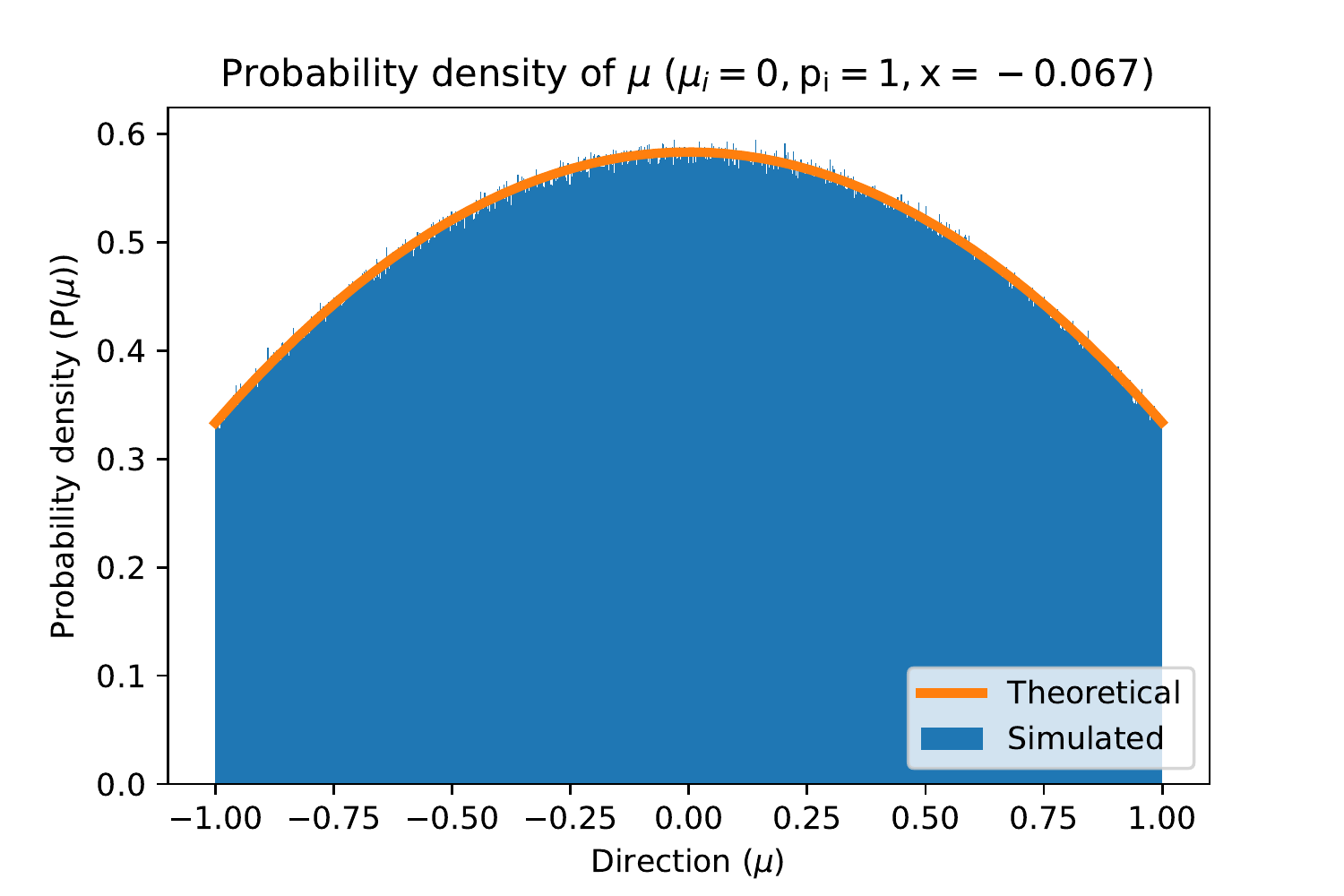}
\includegraphics[width=75mm]{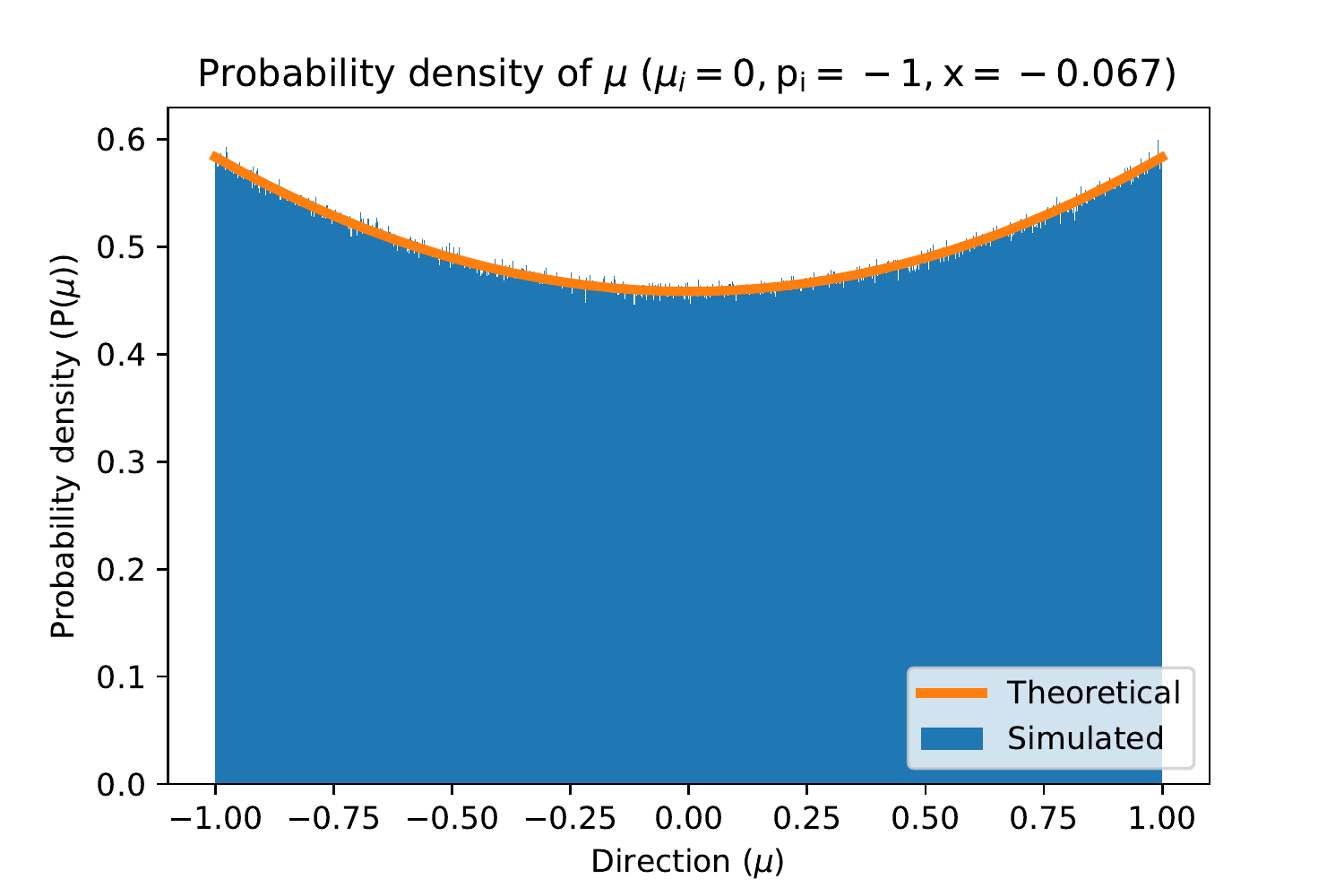}
\caption{\large Simulated and fitted results for the probability density of a photon being emitted in some direction $\mu$. All photons are initially traveling along $\mu_{\rm i} = 0$. The initial polarization values differ in the plots, and the figures represent the different x, which also represent the scattering type. \label{fig:distribution}}
\centering
\end{figure}

Additional tests for the scattering process were done to ensure given an initial direction and polarization, the probability distribution of scattering direction matched theoretical expectations. For initial $\mu_{\rm i} = 0$ and polarization $p_{\rm i} = 0$ and large $x$ (which results in pure dipole scattering, $E_{1}=1$), we see from Eq.~(\ref{equ:3.15}) that
\begin{equation}
    {\rm P}(\boldsymbol{n},{\rm NS}|\boldsymbol{n^{\prime}},p^{\prime}) =  \frac{3}{16\pi}(\cos^2\psi_{\rm NS,NS}+\cos^2\psi_{\rm NS,EW}),
\end{equation}
where $\cos\psi_{NS,NS} = \sin\theta$ and $\cos\psi_{\rm NS,EW} = -\cos\theta \sin\zeta$. 
Using Eq.~(\ref{equ.3.14}), we know 
\begin{equation}
    {\rm P}(\boldsymbol{n},{\rm EW}|\boldsymbol{n^{\prime}},p^{\prime}) =  \frac{3}{16\pi}[\rm cos^2\psi_{\rm EW,NS}+cos^2\psi_{\rm EW,EW}]
\end{equation}
but $\cos\psi_{\rm EW,EW} = 0$, so we see after integrating over all $\zeta$ from 0 to $2\pi$ that
\begin{equation}
    {\rm P}(\boldsymbol{n}|\boldsymbol{n^{\prime}=0},p^{\prime}=0) = \frac{3}{16}(3-\mu^2).
\end{equation}
This matched our simulation results as seen in Fig.~\ref{fig:distribution}. 
Using similar arguments for the isotropic case ($ E_1 = 0$), we can show that ${\rm P}(\boldsymbol{n}|\boldsymbol{n^{\prime}=0},p^{\prime}=0) = \frac12$. For small $x$ case, the scattering type is mixed with about 2/3 pure isotropic and 1/3 pure dipole, thus, when the initial $\mu_{\rm i} = 0$ and initial polarization $p_{\rm i} = 0$, the probability density is
\begin{equation}
    {\rm P}(\boldsymbol{n}|\boldsymbol{n^{\prime}=0},p^{\prime}=0) = \frac{2}{3} \times \frac{1}{2} + \frac{1}{3} \times \frac{3}{16}(3-\mu^2) = \frac{1}{3} + \frac{3-\mu^2}{16},
\end{equation}
which also matches the results of the simulation. 
Other tests with analytic solutions were run with $p_{\rm i}=1, -1$ with pure dipole and mixed condition, with results shown in Table~\ref{tab:my_label}. All simulated distributions had consistent results with the analytic distributions.

\begin{table}[]
    \centering
    \begin{tabular}{|c|c|c|c|} \hline
        scattering type & $p_{\rm i}=-1$ & $p_{\rm i}=0$ & $p_{\rm i}=1$\\
        \hline
        mixed ($E_1=\frac13$; Doppler core) & $\frac13 + \frac18(1 +\mu^2)$ & $\frac13 + \frac1{16}(3 - \mu^2)$ & $\frac13 + \frac14(1 - \mu^2)$\\
        pure dipole ($E_1=1$; damping wing) & $\frac38(1 + \mu^2)$ & $\frac3{16}(3 - \mu^2)$ & $\frac34(1 - \mu^2)$
\\ \hline
    \end{tabular}
    \caption{Analytical results for the probability density of a photon being emitted in some direction $\mu$. All photons are initially traveling along $\mu_{\rm i} = 0$. The initial polarization values differ in the columns, and the rows represent the different scattering law possibilities.}
    \label{tab:my_label}
\end{table}

\subsection{Optically Thin Limit}
\label{thinL}

\begin{figure}
\centering
\subfloat[]{\includegraphics[width=75mm]{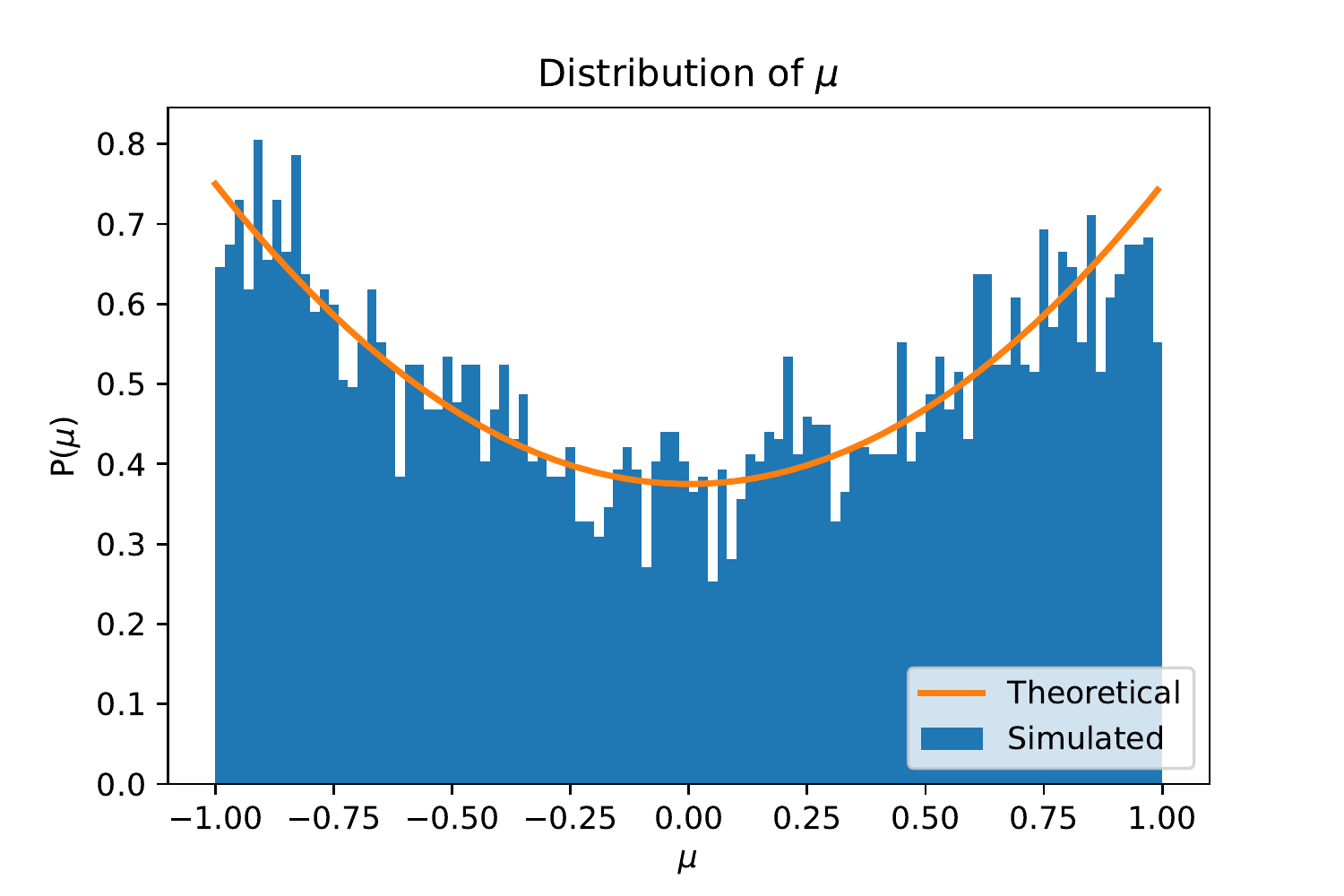}
\label{fig:opticalthin}}
\subfloat[]{\includegraphics[width=75mm]{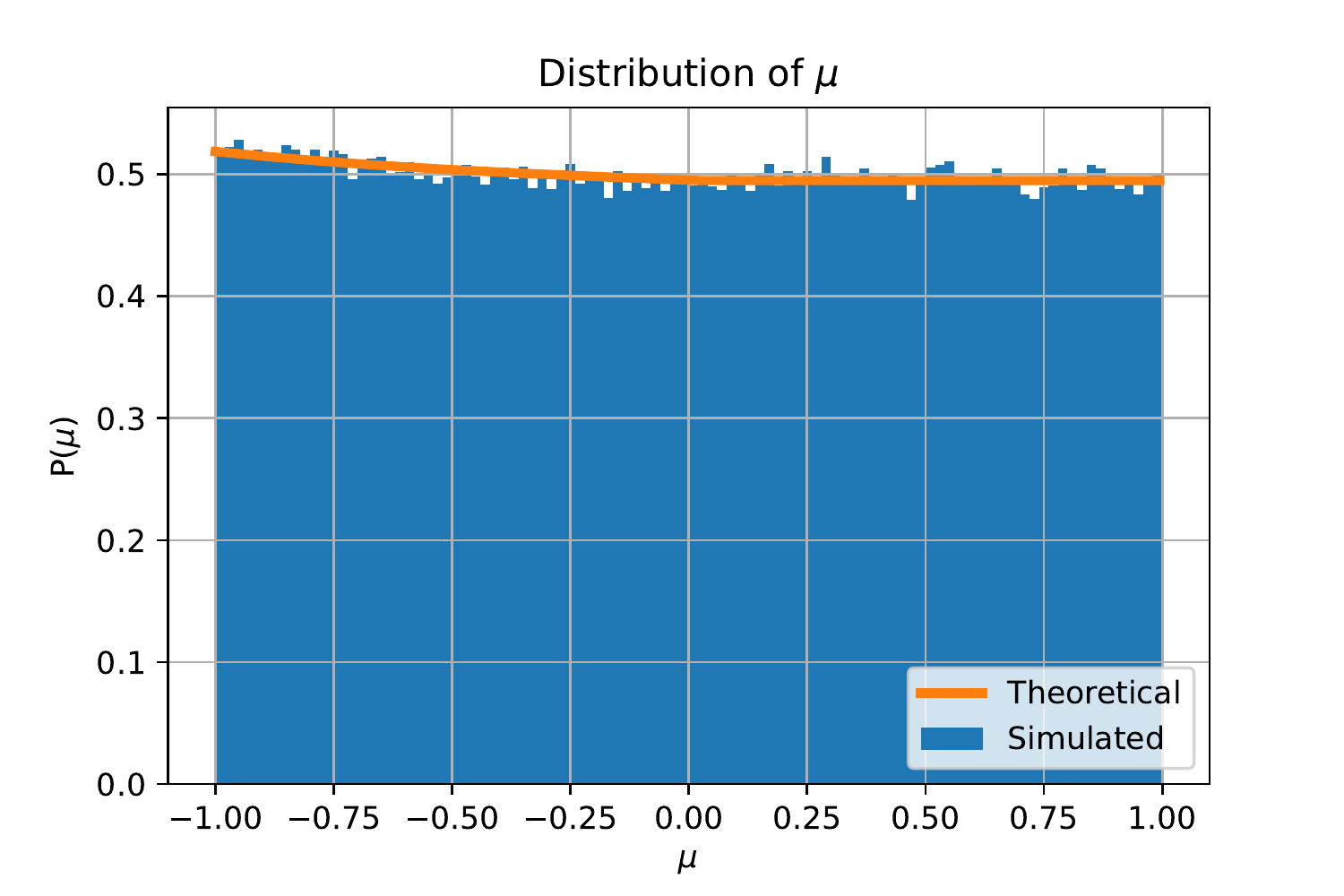}
\label{fig:opticalthick}}
\caption{\large Distribution of scattering direction for optical thin case (a) and optical thick case (b).}
\centering
\end{figure}

To test scattering along the front, we inject photons into the neutral side with $n_{\rm HI} = 1.37 \times 10^{-5}\, {\rm cm}^{-3}$ (neutral fraction of hydrogen is 0.999) in the damping wings. Here we set the photons initially to $p=0$ (unpolarized) and $\mu = 1$ (propagating directly toward the neutral side of the front) with $\Delta \nu = -8.8 \times 10^{12}$ Hz. Then we can calculated the optical depth before the photon exits the frequency grid by taking the Sobolev optical depth \cite{1960mes..book.....S} and multiplying by the fraction of the line profile swept out (the damping wing profile is $\propto\Delta\nu^{-2}$ and can be analytically integrated):
\begin{equation}
    \tau = \frac{3 A^2 n_{\rm HI} \lambda_{\rm center}^3}{32 \pi^3 H} \left(\frac{1}{|\Delta \nu_{\rm ini}|} - \frac{1}{|\Delta \nu_{\rm limit}|}\right).
\end{equation}
For our case, we find $\tau = 0.0555$; then the probability of not scattering is $e^{-\tau}$ = 0.9460 and the probability of scattering is 0.0540. We run $10^5$ photons and get 5340 photons scattering, the probability of scattering is $0.0534\pm 0.0007$ ($1\sigma$ binomial error), which is consistent with expectations.
For the scattering photons, we expect pure dipole scattering since we are in the damping wing, and we should get ${\rm P}(\mu) = \frac38(1+\mu^2)$. Figure~$\ref{fig:opticalthin}$ shows the distribution of scattering photons and corresponds to this expectation.

\subsection{Optically Thick Limit}

Finally, we want to test the optical thick limit. Here we set all other settings the same as optical thin case, but $\Delta \nu = 8.8 \times 10^{12}$ Hz. Since the photons start from the blue side of the line and then redshift toward the line center, we expect nearly all of the photons to scatter. Most of the photons will scatter many times in the neutral region and isotropize, so they approach a uniform distribution of $\mu$ with $\rm P(\mu) = 1/2$. It will have a slight excess of photon with $\mu\approx -1$ as they may be single scatterings that go back and enter the ionized region before they redshift into the line center. We run $5 \times 10^5$ photons in this case to reduce the noise to signal ratio, and after excluding the single scatterings case, we get the result shown in Figure~\ref{fig:opticalthick}, which is corresponding to our expectation.
In this test, the theoretical distribution would be
\begin{equation}
    P(\mu) = \frac12(1-P_{\rm all}) + \begin{cases}
P_{\rm 1 \, scatter}(\mu) &  \mu < 0 \\ 0 & \mu \ge 0,
\end{cases}
\end{equation}
where
\begin{equation}
    P_{\rm 1 \, scatter}(\mu) = \frac38(1+\mu^2)\,\tau\,\int_0^{\frac{1}{1-1/\mu}} e^{-\tau\,(1-1/\mu)}\,\frac{x}{[1-(1-1/\mu)\,x]\,(1-x)} dx,
\end{equation}
we have defined
$P_{\rm all} = \int_{-1}^0 P_{\rm 1\,scatter}(\mu)\, d\mu$,
and $\tau$ is the same as in Eq.~(\ref{thinL}).

\bibliographystyle{JHEP.bst}
\bibliography{main.bib}

\end{document}